\documentstyle[12pt,epsfig,graphics,cite]{article}
\begin{document}\bibliographystyle{plain}\begin{titlepage}
\renewcommand{\thefootnote}{\fnsymbol{footnote}}\hfill
\begin{tabular}{l}HEPHY-PUB 843/07\\November 2007\end{tabular}\\[.5cm]
\Large\begin{center}{\bf STABILITY IN THE INSTANTANEOUS
BETHE--SALPETER FORMALISM: HARMONIC-OSCILLATOR REDUCED SALPETER
EQUATION}\\[1cm]\large{\bf Zhi-Feng LI}\\[.3cm]\normalsize Faculty
of Physics, University of Vienna,\\Boltzmanngasse 5, A-1090
Vienna, Austria\\[1cm]\large{\bf Wolfgang
LUCHA\footnote[1]{\normalsize\ {\em E-mail address\/}:
wolfgang.lucha@oeaw.ac.at}}\\[.3cm]\normalsize Institute for High
Energy Physics,\\Austrian Academy of Sciences,\\Nikolsdorfergasse
18, A-1050 Vienna, Austria\\[1cm]\large{\bf Franz
F.~SCH\"OBERL\footnote[2]{\normalsize\ {\em E-mail address\/}:
franz.schoeberl@univie.ac.at}}\\[.3cm]\normalsize Faculty of
Physics, University of Vienna,\\Boltzmanngasse 5, A-1090 Vienna,
Austria\vfill{\normalsize\bf Abstract}\end{center}\normalsize A
popular three-dimensional reduction of the Bethe--Salpeter
formalism for the description of bound states in quantum field
theory is the Salpeter equation, derived by assuming~both
instantaneous interactions and free propagation of all bound-state
constituents. Numerical (variational) studies of the Salpeter
equation with confining interaction, however, observed specific
instabilities of the solutions, likely related to the Klein
paradox and rendering~(part of the) bound states unstable. An
analytic investigation of the problem by a comprehensive spectral
analysis is feasible for the reduced Salpeter equation with only
harmonic-oscillator confining interactions. There we are able to
prove rigorously that the bound-state solutions correspond to real
discrete spectra bounded from below and are thus free of all
instabilities.\vspace{.5cm}

\renewcommand{\thefootnote}{\arabic{footnote}}\end{titlepage}

\section{Introduction}The most widely explored three-dimensional
reduction of the Bethe--Salpeter formalism \cite{BSE} for the
description of bound states within quantum field theory is the
Salpeter equation \cite{SE}. The Salpeter equation controls the
Salpeter amplitude, which in momentum space encodes the
distribution of relative momenta of all bound-state constituents.
In elementary particle physics its application to quantum
electrodynamics (QED) and quantum chromodynamics (QCD) has met
considerable success. In particular, within the latter realm it
has evolved to a well-established standard tool for describing
from first principles hadrons as bound states of quarks, confined
by the strong interactions. Surprisingly or not, however, the
solutions of Salpeter's equation with {\em confining\/}
interactions have been {\em numerically\/} shown to develop~for
some Lorentz structures of the Bethe--Salpeter kernel representing
all interactions between the bound-state constituents
instabilities which cause states expected to be stable to decay.

In contrast, the {\em reduced\/} Salpeter equation
\cite{Henriques76,Jacobs87,Gara89,Gara90,Lucha92C}, derived from
the full Salpeter equation by neglecting some of the interaction
terms (Sec.~\ref{Sec:RIBSE}), offers the chance to study the
question of stability {\em analytically}. Such instabilities
should arise first for pseudoscalar states (Sec.~\ref{Sec:PSBS});
there, stripping off all angular variables simplifies
\cite{Lagae92a,Lagae92b,Olsson95,Olsson96,
Lucha00:IBSEm=0,Lucha00:IBSE-C4,Lucha00:IBSEnzm,Lucha01:IBSEIAS}
the reduced Salpeter equation to a single integral equation
(Sec.~\ref{Sec:REE}). Harmonic-oscillator confining interactions
(Sec.~\ref{Sec:HOI}) have a big advantage: In momentum space, they
are represented by a simple Laplacian,~converting thus our
integral to differential equations (Sec.~\ref{Sec:ODE}). Their
analysis proves that, for all famous Lorentz structures, including
one proposed by B\"ohm, Joos, and Krammer (BJK hereafter)
\cite{BJK73}, studied in Ref.~\cite{Gross91} and more recently
used, among others, by a group in Bonn
\cite{Koll00,Ricken00,Merten01,Ricken03}, any bound-state solution
is related to a real (Sec.~\ref{Sec:SA}) and discrete
(Sec.~\ref{Sec:SAS}) energy eigenvalue, and thus stable. Similar
considerations can be applied to the full Salpeter equation
(Sec.~\ref{Sec:GFSE}).

\section{Bethe--Salpeter Formalism in Instantaneous
Limit}\label{Sec:RIBSE}We are interested in instantaneous
approximations to Bethe--Salpeter equations describing bound
states composed of a fermion and an antifermion. Let these
constituents, denoted~by $i=1,2,$ carry the momenta $p_1$ and
$p_2,$ which will enter also in terms of the total momentum
$P\equiv p_1+p_2$ and, for a real parameter $\eta$ or $\zeta,$ the
relative momentum $p\equiv p_1-\eta\,P\equiv\zeta\,P-p_2.$

\subsection{Salpeter equation}\label{Sec:SE}Any derivation of
the Salpeter equation \cite{SE} as one of a variety of possible
three-dimensional reductions of the Bethe--Salpeter formalism is
based on just two fundamental assumptions: First, one obtains the
{\em instantaneous Bethe--Salpeter equation\/} if all interactions
between~the bound-state constituents are instantaneous in the
center-of-momentum frame of the bound state. In the
Bethe--Salpeter equation \cite{BSE} all interactions between
bound-state constituents are encoded in their integral kernel,
$K.$ The instantaneous approximation then implies that $K$ depends
only on the spatial components of the relative momenta involved:
$K=K(\mbox{\boldmath{$p$}},\mbox{\boldmath{$q$}}).$ Second, one
arrives at the {\em Salpeter equation\/} if every bound-state
constituent propagates as free particle with effective mass, $m.$
Salpeter's equation adopts the free fermion
propagator$$S_0(p,m)=\frac{{\rm i}}{\not\!p-m+{\rm
i}\,\varepsilon}\equiv{\rm i}\,\frac{\not\!p+m}{p^2-m^2+{\rm
i}\,\varepsilon}\ ,\quad\varepsilon\downarrow0\ .$$

Let us present those aspects of the Salpeter equation which will
be relevant for stability. We introduce the one-particle energy
$E_i(\mbox{\boldmath{$p$}}),$ the one-particle Dirac Hamiltonian
$H_i(\mbox{\boldmath{$p$}}),$~and the energy projection operators
$\Lambda_i^\pm(\mbox{\boldmath{$p$}})$ for positive or negative
energy of particle
$i=1,2$~by\begin{eqnarray*}E_i(\mbox{\boldmath{$p$}})&\equiv&
\sqrt{\mbox{\boldmath{$p$}}^2+m_i^2}\ ,\quad i=1,2\
,\\[1ex]H_i(\mbox{\boldmath{$p$}})&\equiv&
\gamma_0\,(\mbox{\boldmath{$\gamma$}}\cdot\mbox{\boldmath{$p$}}+m_i)\
,\quad i=1,2\ ,\\[1ex]
\Lambda_i^\pm(\mbox{\boldmath{$p$}})&\equiv&\frac{E_i(\mbox{\boldmath{$p$}})\pm
H_i(\mbox{\boldmath{$p$}})}{2\,E_i(\mbox{\boldmath{$p$}})}\ ,\quad
i=1,2\ ;\end{eqnarray*}related, for $i=1,2,$ by
$H_i^2(\mbox{\boldmath{$p$}})=E_i^2(\mbox{\boldmath{$p$}})$ and
$\Lambda_i^\pm(\mbox{\boldmath{$p$}})\,H_i(\mbox{\boldmath{$p$}})
=H_i(\mbox{\boldmath{$p$}})\,\Lambda_i^\pm(\mbox{\boldmath{$p$}})
=\pm\,E_i(\mbox{\boldmath{$p$}})\,\Lambda_i^\pm(\mbox{\boldmath{$p$}}),$
all operators $\Lambda_i^\pm(\mbox{\boldmath{$p$}})$ fulfilling
$\Lambda_i^\pm(\mbox{\boldmath{$p$}})\,\Lambda_i^\pm(\mbox{\boldmath{$p$}})
=\Lambda_i^\pm(\mbox{\boldmath{$p$}}),$
$\Lambda_i^\pm(\mbox{\boldmath{$p$}})\,\Lambda_i^\mp(\mbox{\boldmath{$p$}})
=0,$
$\Lambda_i^+(\mbox{\boldmath{$p$}})+\Lambda_i^-(\mbox{\boldmath{$p$}})=1.$
In terms of the above abbreviations, the Salpeter equation
\cite{SE}, for bound states of a~fermion (of mass $m_1$ and
momentum $\mbox{\boldmath{$p$}}_1$) and an antifermion (of mass
$m_2$ and momentum $\mbox{\boldmath{$p$}}_2$),~reads
\begin{eqnarray}\Phi(\mbox{\boldmath{$p$}})&=&\int\frac{{\rm
d}^3q}{(2\pi)^3}\left(\frac{\Lambda_1^+(\mbox{\boldmath{$p$}}_1)\,\gamma_0\,
[K(\mbox{\boldmath{$p$}},\mbox{\boldmath{$q$}})\,
\Phi(\mbox{\boldmath{$q$}})]\,\gamma_0\,\Lambda_2^-(\mbox{\boldmath{$p$}}_2)}
{P_0-E_1(\mbox{\boldmath{$p$}}_1)-E_2(\mbox{\boldmath{$p$}}_2)}\right.\nonumber\\[1ex]
&&\hspace{7.88ex}\left.-\frac{\Lambda_1^-(\mbox{\boldmath{$p$}}_1)\,\gamma_0\,
[K(\mbox{\boldmath{$p$}},\mbox{\boldmath{$q$}})\,
\Phi(\mbox{\boldmath{$q$}})]\,\gamma_0\,\Lambda_2^+(\mbox{\boldmath{$p$}}_2)}
{P_0+E_1(\mbox{\boldmath{$p$}}_1)+E_2(\mbox{\boldmath{$p$}}_2)}\right).
\label{Eq:SE}\end{eqnarray}Accordingly, every solution
$\Phi(\mbox{\boldmath{$p$}})$ of the Salpeter equation has to
satisfy the two
constraints$$\Lambda_1^+(\mbox{\boldmath{$p$}}_1)\,
\Phi(\mbox{\boldmath{$p$}})\,\Lambda_2^+(\mbox{\boldmath{$p$}}_2)
=\Lambda_1^-(\mbox{\boldmath{$p$}}_1)\,\Phi(\mbox{\boldmath{$p$}})\,
\Lambda_2^-(\mbox{\boldmath{$p$}}_2)=0\ ,$$which, by considering
their sum or difference, prove to be equivalent to a single
constraint:\begin{equation}\frac{H_1(\mbox{\boldmath{$p$}}_1)}{E_1(\mbox{\boldmath{$p$}}_1)}
\,\Phi(\mbox{\boldmath{$p$}})+\Phi(\mbox{\boldmath{$p$}})\,
\frac{H_2(\mbox{\boldmath{$p$}}_2)}{E_2(\mbox{\boldmath{$p$}}_2)}=0\label{Eq:SEC}\end{equation}or,
equivalently,$$\Phi(\mbox{\boldmath{$p$}})
+\frac{H_1(\mbox{\boldmath{$p$}}_1)\,\Phi(\mbox{\boldmath{$p$}})
\,H_2(\mbox{\boldmath{$p$}}_2)}{E_1(\mbox{\boldmath{$p$}}_1)\,E_2(\mbox{\boldmath{$p$}}_2)}=0\
.$$With the help of the decomposition of unity in terms of
projection operators $\Lambda_i^\pm(\mbox{\boldmath{$p$}}),$ that
is,
$1\otimes1=\Lambda_1^+(\mbox{\boldmath{$p$}}_1)\otimes\Lambda_2^+(\mbox{\boldmath{$p$}}_2)+
\Lambda_1^+(\mbox{\boldmath{$p$}}_1)\otimes\Lambda_2^-(\mbox{\boldmath{$p$}}_2)+
\Lambda_1^-(\mbox{\boldmath{$p$}}_1)\otimes\Lambda_2^+(\mbox{\boldmath{$p$}}_2)+
\Lambda_1^-(\mbox{\boldmath{$p$}}_1)\otimes\Lambda_2^-(\mbox{\boldmath{$p$}}_2),$
a Salpeter amplitude $\Phi(\mbox{\boldmath{$p$}})$ may be, in
general, decomposed into the components
$\Lambda_1^\pm(\mbox{\boldmath{$p$}}_1)\,\Phi(\mbox{\boldmath{$p$}})\,\Lambda_2^\pm(\mbox{\boldmath{$p$}}_2)$:
\begin{eqnarray*}\Phi(\mbox{\boldmath{$p$}})&=&
\Lambda_1^+(\mbox{\boldmath{$p$}}_1)\,\Phi(\mbox{\boldmath{$p$}})\,\Lambda_2^+(\mbox{\boldmath{$p$}}_2)+
\Lambda_1^+(\mbox{\boldmath{$p$}}_1)\,\Phi(\mbox{\boldmath{$p$}})\,\Lambda_2^-(\mbox{\boldmath{$p$}}_2)\\&+&
\Lambda_1^-(\mbox{\boldmath{$p$}}_1)\,\Phi(\mbox{\boldmath{$p$}})\,\Lambda_2^+(\mbox{\boldmath{$p$}}_2)+
\Lambda_1^-(\mbox{\boldmath{$p$}}_1)\,\Phi(\mbox{\boldmath{$p$}})\,\Lambda_2^-(\mbox{\boldmath{$p$}}_2)\
.\end{eqnarray*}The above constraint(s) on the solutions
$\Phi(\mbox{\boldmath{$p$}})$ of the Salpeter equation
(\ref{Eq:SE}), arising from its specific projector structure,
halve, in fact, the number of independent
components~of~$\Phi(\mbox{\boldmath{$p$}})$:\begin{equation}\Phi(\mbox{\boldmath{$p$}})=
\Lambda_1^+(\mbox{\boldmath{$p$}}_1)\,\Phi(\mbox{\boldmath{$p$}})\,\Lambda_2^-(\mbox{\boldmath{$p$}}_2)+
\Lambda_1^-(\mbox{\boldmath{$p$}}_1)\,\Phi(\mbox{\boldmath{$p$}})\,\Lambda_2^+(\mbox{\boldmath{$p$}}_2)\
.\label{Eq:SEsoln}\end{equation}Assuming the Lorentz structures of
the effective couplings of the bound fermions $i=1,2$ to the
corresponding interaction potentials to be identical, the
Bethe--Salpeter kernel
$K(\mbox{\boldmath{$p$}},\mbox{\boldmath{$q$}})$ is, quite
generally, the sum of products of a tensor product
$\Gamma\otimes\Gamma$ of generic Dirac matrices $\Gamma$ and a
Lorentz-scalar associated interaction function
$V_\Gamma(\mbox{\boldmath{$p$}},\mbox{\boldmath{$q$}})$:
$K(\mbox{\boldmath{$p$}},\mbox{\boldmath{$q$}})=\sum_\Gamma
V_\Gamma(\mbox{\boldmath{$p$}},\mbox{\boldmath{$q$}})\,\Gamma\otimes\Gamma.$
More precisely, the action of the kernel
$K(\mbox{\boldmath{$p$}},\mbox{\boldmath{$q$}})$ on the Salpeter
amplitude $\Phi(\mbox{\boldmath{$p$}})$ is given~by
$$[K(\mbox{\boldmath{$p$}},\mbox{\boldmath{$q$}})\,\Phi(\mbox{\boldmath{$q$}})]=\sum_\Gamma
V_\Gamma(\mbox{\boldmath{$p$}},\mbox{\boldmath{$q$}})\,\Gamma\,\Phi(\mbox{\boldmath{$q$}})\,\Gamma\
.$$

\subsection{Reduced Salpeter equation}\label{Sec:RSE}The
projection operators $\Lambda_i^\pm(\mbox{\boldmath{$p$}}),$
$i=1,2,$ for positive or negative energy satisfy the~identity
$$\left[\Lambda_i^\pm(\mbox{\boldmath{$p$}})\right]^{\rm c}\equiv
\left[C^{-1}\,\Lambda_i^\pm(\mbox{\boldmath{$p$}})\,C\right]^{\rm
T}=\Lambda_i^\mp(\mbox{\boldmath{$p$}})\ ,\quad i=1,2\ ,$$where
$C$ denotes the usual Dirac-space charge-conjugation matrix.
Accordingly, the second term on the right-hand side of the
Salpeter equation, Eq.~(\ref{Eq:SE}), is that interaction term
which is related to the {\em negative-energy components\/}
$\Lambda_1^-(\mbox{\boldmath{$p$}}_1)\,\Phi(\mbox{\boldmath{$p$}})\,
\Lambda_2^+(\mbox{\boldmath{$p$}}_2)=\Lambda_1^-(\mbox{\boldmath{$p$}}_1)\,
\Phi(\mbox{\boldmath{$p$}})\,[\Lambda_2^-(\mbox{\boldmath{$p$}}_2)]^{\rm
c}$ of the Salpeter amplitude $\Phi(\mbox{\boldmath{$p$}}).$
Assuming that this term's contribution may be reasonably ignored
relative to that of the first term yields the so-called reduced
Salpeter equation
\cite{Henriques76,Jacobs87,Gara89,Gara90,Lucha92C}\begin{equation}
\left[P_0-E_1(\mbox{\boldmath{$p$}}_1)-E_2(\mbox{\boldmath{$p$}}_2)\right]
\Phi(\mbox{\boldmath{$p$}})=\int\frac{{\rm
d}^3q}{(2\pi)^3}\,\Lambda_1^+(\mbox{\boldmath{$p$}}_1)\,\gamma_0\,
[K(\mbox{\boldmath{$p$}},\mbox{\boldmath{$q$}})\,
\Phi(\mbox{\boldmath{$q$}})]\,\gamma_0\,\Lambda_2^-(\mbox{\boldmath{$p$}}_2)\
.\label{Eq:RSE}\end{equation}This neglect might be justifiable for
nonrelativistic and weakly bound systems composed of heavy
constituents, such that, on the average,
$P_0-E_1(\mbox{\boldmath{$p$}}_1)-E_2(\mbox{\boldmath{$p$}}_2)\ll
P_0+E_1(\mbox{\boldmath{$p$}}_1)+E_2(\mbox{\boldmath{$p$}}_2).$
More rigorously, one would like to be sure, at least, that, for
appropriate {\em expectation values},
$$\frac{1}{P_0+E_1(\mbox{\boldmath{$p$}}_1)+E_2(\mbox{\boldmath{$p$}}_2)}\ll
\frac{1}{P_0-E_1(\mbox{\boldmath{$p$}}_1)-E_2(\mbox{\boldmath{$p$}}_2)}\
.$$Formally, this reduction of the (full) Salpeter equation
(\ref{Eq:SE}) to the reduced Salpeter equation (\ref{Eq:RSE}) can
be accomplished by subjecting the Salpeter amplitude
$\Phi(\mbox{\boldmath{$p$}})$ to {\em any\/} of the equalities
$$\Lambda_1^-(\mbox{\boldmath{$p$}}_1)\,\Phi(\mbox{\boldmath{$p$}})=0\
,\quad\Phi(\mbox{\boldmath{$p$}})\,\Lambda_2^+(\mbox{\boldmath{$p$}}_2)=0\
,$$or,
equivalently,$$H_1(\mbox{\boldmath{$p$}}_1)\,\Phi(\mbox{\boldmath{$p$}})=
E_1(\mbox{\boldmath{$p$}}_1)\,\Phi(\mbox{\boldmath{$p$}})\ ,\quad
\Phi(\mbox{\boldmath{$p$}})\,H_2(\mbox{\boldmath{$p$}}_2)=-
E_2(\mbox{\boldmath{$p$}}_2)\,\Phi(\mbox{\boldmath{$p$}})\
;$$because of the particular projector structure of the resulting
reduced Salpeter equation (\ref{Eq:RSE}), imposition of one of
these two constraints automatically implies the other for the
solutions. These equalities entail the constraints
$\Lambda_1^+(\mbox{\boldmath{$p$}}_1)\,
\Phi(\mbox{\boldmath{$p$}})\,\Lambda_2^+(\mbox{\boldmath{$p$}}_2)=
\Lambda_1^-(\mbox{\boldmath{$p$}}_1)\,\Phi(\mbox{\boldmath{$p$}})\,
\Lambda_2^-(\mbox{\boldmath{$p$}}_2)=0$ also satisfied by each
solution of the Salpeter equation (\ref{Eq:SE}), as well as, in
addition, the constraint $\Lambda_1^-(\mbox{\boldmath{$p$}}_1)\,
\Phi(\mbox{\boldmath{$p$}})\,\Lambda_2^+(\mbox{\boldmath{$p$}}_2)=0.$
As trivial consequence, any solution $\Phi(\mbox{\boldmath{$p$}})$
of the reduced Salpeter equation (\ref{Eq:RSE}) is necessarily of
the unique component structure $\Phi(\mbox{\boldmath{$p$}})=
\Lambda_1^+(\mbox{\boldmath{$p$}}_1)\,\Phi(\mbox{\boldmath{$p$}})\,
\Lambda_2^-(\mbox{\boldmath{$p$}}_2).$

\section{Pseudoscalar Bound States}\label{Sec:PSBS}The states
easiest to investigate are bound states composed of a fermion and
the associated antifermion, which guarantees a well-defined
behaviour with respect to charge conjugation. The masses of a
particle and the corresponding antiparticle are, of course,
identical, that is, $m_1=m_2=m.$ Now, in the {\em
center-of-momentum\/} (or rest) {\em frame\/} of the two-particle
system under study, defined by $\mbox{\boldmath{$P$}}={\bf 0},$
which implies $\mbox{\boldmath{$p$}}=
\mbox{\boldmath{$p$}}_1=-\mbox{\boldmath{$p$}}_2,$ the time
component $P_0$ of the total momentum $P$ reduces to the
bound-state mass eigenvalue $M,$ i.e., $P_0=M.$ All indices
$i=1,2,$ distinguishing the two bound-state constituents, may then
be dropped throughout the following analysis:
$E_1(\mbox{\boldmath{$p$}})=E_2(\mbox{\boldmath{$p$}})=
E(\mbox{\boldmath{$p$}})=E(p)\equiv\sqrt{p^2+m^2},$ where~$p\equiv
|\mbox{\boldmath{$p$}}|\equiv\sqrt{\mbox{\boldmath{$p$}}^2};$ and,
{\em mutatis mutandis}, for the Hamiltonians
$H_i(\mbox{\boldmath{$p$}})$ and energy projection operators
$\Lambda_i^\pm(\mbox{\boldmath{$p$}}).$

For the sake of definiteness we focus, in what follows, to the
case of fermion--antifermion bound states of total spin $J,$
parity $P=(-1)^{J+1},$ and charge-conjugation quantum number
$C=(-1)^J$ (which entails $C\,P=-1$ for all $J$). In usual
spectroscopic notation, these~states are denoted by $n\,^1J_J$
($n=1,2,3,\dots$). As the perhaps simplest example within this
context, precisely these physical systems have been studied rather
frequently (cf., e.g.,
Refs.~\cite{Lagae92a,Lagae92b,Olsson95,Olsson96,
Lucha00:IBSEm=0,Lucha00:IBSE-C4,Lucha00:IBSEnzm,Lucha01:IBSEIAS}).

More specifically, we consider, for reasons of simplicity, bound
states of spin $J=0,$ that is, pseudoscalar bound states, with
spin-parity-charge conjugation assignment $J^{PC}=0^{-+},$
spectroscopically called ${}^1{\rm S}_0.$ In nature such systems
are realized and observed, for instance, in the realm of
quark--antiquark bound states, in form of the pion and its radial
excitations.

In the case of the {\em Salpeter equation\/} (\ref{Eq:SE}), as a
consequence of the constraints discussed in Subsec.~\ref{Sec:SE}
the most general expansion of the Salpeter amplitude
$\Phi(\mbox{\boldmath{$p$}}),$ over a complete set of Dirac
matrices, involves not the full 16 but only eight {\em
independent\/} Salpeter components. For the description of
${}^1J_J$ states, only two of the latter, called
$\phi_1(\mbox{\boldmath{$p$}})$ and
$\phi_2(\mbox{\boldmath{$p$}}),$ are relevant. With our notation
for one-particle energy $E(\mbox{\boldmath{$p$}})$ and Dirac
Hamiltonian $H(\mbox{\boldmath{$p$}})$ introduced in
Subsec.~\ref{Sec:SE}, full ${}^1J_J$ Salpeter amplitudes
$\Phi(\mbox{\boldmath{$p$}})$ for fermion and antifermion of equal
mass $m$ and internal momentum $\mbox{\boldmath{$p$}}$ thus read,
in the center-of-momentum frame~of~the~bound~state,
$$\Phi(\mbox{\boldmath{$p$}})=\left[\phi_1(\mbox{\boldmath{$p$}})\,
\frac{H(\mbox{\boldmath{$p$}})}{E(\mbox{\boldmath{$p$}})}
+\phi_2(\mbox{\boldmath{$p$}})\right]\gamma_5\ .$$

In the case of the {\em reduced Salpeter equation\/}
(\ref{Eq:RSE}), the one additional, two-faced constraint analyzed
in Subsec.~\ref{Sec:RSE} entails, possibly taking into account
$\gamma_5\,H(-\mbox{\boldmath{$p$}})=-H(\mbox{\boldmath{$p$}})\,
\gamma_5,$ for the two independent Salpeter components
$\phi_1(\mbox{\boldmath{$p$}})$ and
$\phi_2(\mbox{\boldmath{$p$}})$ in the ${}^1J_J$ Salpeter
amplitude~$\Phi(\mbox{\boldmath{$p$}}),$
$\phi_1(\mbox{\boldmath{$p$}})=\phi_2(\mbox{\boldmath{$p$}})
\equiv\phi(\mbox{\boldmath{$p$}}).$ Consequently, the generic
${}^1J_J$ reduced-Salpeter solutions $\Phi(\mbox{\boldmath{$p$}})$
read$$\Phi(\mbox{\boldmath{$p$}})=\phi(\mbox{\boldmath{$p$}})\,
\frac{H(\mbox{\boldmath{$p$}})+E(\mbox{\boldmath{$p$}})}
{E(\mbox{\boldmath{$p$}})}\,\gamma_5\equiv2\,
\phi(\mbox{\boldmath{$p$}})\,\Lambda^+(\mbox{\boldmath{$p$}})\,
\gamma_5\ .$$

\section{Radial Eigenvalue Equations}\label{Sec:REE}For any
(instantaneous) Bethe--Salpeter interaction kernel
$K(\mbox{\boldmath{$p$}},\mbox{\boldmath{$q$}})$ of {\em
convolution type}, i.e.,
$K(\mbox{\boldmath{$p$}},\mbox{\boldmath{$q$}})
=K(\mbox{\boldmath{$p$}}-\mbox{\boldmath{$q$}})$ and therefore
$V_\Gamma(\mbox{\boldmath{$p$}},\mbox{\boldmath{$q$}})
=V_\Gamma(\mbox{\boldmath{$p$}}-\mbox{\boldmath{$q$}}),$ by
factorizing off all dependence on angular variables encoded in
corresponding (vector) spherical harmonics both the Salpeter
equation (\ref{Eq:SE}) \cite{Lagae92a,Olsson95} and its reduced
version (\ref{Eq:RSE}) \cite{Olsson96} can be converted into
equivalent~systems of coupled equations for the radial factors of
all relevant independent Salpeter components. For a fixed Lorentz
structure of the kernel, the interactions experienced by the
bound-state constituents enter in such set of equations in the
form of Fourier--Bessel transforms $V_L(p,q)$ ($L=0,1,2,\dots$) of
some spherically symmetric static potential $V(r)$ in
configuration space:$$V_L(p,q)\equiv 8\pi\int\limits_0^\infty{\rm
d}r\,r^2\,j_L(p\,r)\,j_L(q\,r)\,V(r)\ ,\quad L=0,1,2,\dots\ ,$$
where $j_n(z),$ for $n=0,\pm 1,\pm2,\dots,$ label the spherical
Bessel functions of the first~kind~\cite{Abramowitz}.

According to Sec.~\ref{Sec:PSBS}, for pseudoscalar
fermion--antifermion bound states each solution of the reduced
Salpeter equation (\ref{Eq:RSE}) involves only one independent
Salpeter component, $\phi(p).$ Consequently, the aforementioned
system of radial equations collapses to a single~equation. The
application of the radial reduction to the reduced Salpeter
equation (\ref{Eq:RSE}) then yields the radial eigenvalue
equations, for interactions of Lorentz-scalar Dirac structure,
$\Gamma\otimes\Gamma=1\otimes1,$
$$2\,E(p)\,\phi(p)-\frac{1}{2}\int\limits_0^\infty\frac{{\rm
d}q\,q^2}{(2\pi)^2}\left[\left(1+\frac{m^2}{E(p)\,E(q)}\right)V_0(p,q)
-\frac{p\,q}{E(p)\,E(q)}\,V_1(p,q)\right]\phi(q)=M\,\phi(p)\ ,$$
for interactions of time-component Lorentz-vector Dirac structure,
$\Gamma\otimes\Gamma=\gamma^0\otimes\gamma^0,$
$$2\,E(p)\,\phi(p)+\frac{1}{2}\int\limits_0^\infty\frac{{\rm
d}q\,q^2}{(2\pi)^2}\left[\left(1+\frac{m^2}{E(p)\,E(q)}\right)V_0(p,q)
+\frac{p\,q}{E(p)\,E(q)}\,V_1(p,q)\right]\phi(q)=M\,\phi(p)\ ,$$
for interactions of Lorentz-vector Dirac structure,
$\Gamma\otimes\Gamma=\gamma_\mu\otimes\gamma^\mu,$
$$2\,E(p)\,\phi(p)+\int\limits_0^\infty\frac{{\rm
d}q\,q^2}{(2\pi)^2}\left(2-\frac{m^2}{E(p)\,E(q)}\right)V_0(p,q)\,
\phi(q)=M\,\phi(p)\ ,$$for interactions of Lorentz-pseudoscalar
Dirac structure, $\Gamma\otimes\Gamma=\gamma_5\otimes\gamma_5,$
$$2\,E(p)\,\phi(p)-\frac{1}{2}\int\limits_0^\infty\frac{{\rm
d}q\,q^2}{(2\pi)^2}\left[\left(1-\frac{m^2}{E(p)\,E(q)}\right)V_0(p,q)
-\frac{p\,q}{E(p)\,E(q)}\,V_1(p,q)\right]\phi(q)=M\,\phi(p)\ ,$$
and, for interactions of BJK \cite{BJK73,Gross91} Dirac structure,
$\Gamma\otimes\Gamma=\frac{1}{2}\left(
\gamma_\mu\otimes\gamma^\mu+\gamma_5\otimes\gamma_5-1\otimes1\right),$
$$2\,E(p)\,\phi(p)+\int\limits_0^\infty\frac{{\rm
d}q\,q^2}{(2\pi)^2}\,V_0(p,q)\,\phi(q)=M\,\phi(p)\ .$$For any mass
eigenvalue $M$ such that $M-2\,E(p)\ne0$ the above radial
eigenvalue equations are all of the form of a homogeneous linear
(Fredholm) integral equation of the second~kind.

\section{Harmonic-Oscillator Interaction}\label{Sec:HOI}The reduced
Salpeter equations of Sec.~\ref{Sec:REE} may be discussed to a
large extent analytically by focusing to harmonic-oscillator
interactions described by the configuration-space
potential$$V(r)=a\,r^2\ ,\quad a=a^\ast\ne0\ ,\quad r\equiv
|\mbox{\boldmath{$x$}}|\ .$$For this choice of $V(r)$ we are able
to determine analytically all potential functions $V_L(p,q)$
entering in the radial eigenvalue equations, by taking advantage
of the differential equation \cite{Abramowitz} satisfied by {\em
all\/} spherical Bessel functions, generically called $w_n(z)$
($n=0,\pm1,\pm2,\dots$):$$z^2\,\frac{{\rm d}^2}{{\rm
d}z^2}w_n(z)+2\,z\,\frac{{\rm d}}{{\rm
d}z}w_n(z)+[z^2-n\,(n+1)]\,w_n(z)=0\ .$$Defining, as radial relic
of the Laplacian
$\Delta\equiv\mbox{\boldmath{$\nabla$}}\cdot\mbox{\boldmath{$\nabla$}},$
the second-order differential operators
$$D_p^{(L)}\equiv\frac{{\rm d}^2}{{\rm
d}p^2}+\frac{2}{p}\,\frac{{\rm d}}{{\rm d}p}-\frac{L\,(L+1)}{p^2}\
,\quad L=0,1,2,\dots\ ,$$the spherical Bessel functions of the
first kind, $j_L(p\,r),$ in the potentials $V_L(p,q)$ thus~satisfy
$$D_p^{(L)}j_L(p\,r)=-r^2\,j_L(p\,r)\ ,\quad L=0,1,2,\dots\ .$$
This relation allows to replace the harmonic-oscillator potential
$r^2$ in the potential function $V_L(p,q)$ by the differential
operator $D_p^{(L)}.$ Hence, by means of the ``orthogonality
relations''$$\int\limits_0^\infty{\rm
d}r\,r^2\,j_L(p\,r)\,j_L(q\,r)=\frac{\pi}{2\,p^2}\,\delta(p-q)\
,\quad L=0,1,2,\dots\ ,$$involving Dirac's delta distribution the
potential functions for harmonic oscillators become
\begin{equation}V_L(p,q)=-\frac{(2\pi)^2\,a}{q^2}\,D_p^{(L)}\delta(p-q)\
,\quad L=0,1,2,\dots\ .\label{Eq:HOI}\end{equation}

\section{Ordinary Differential Equations (of Second Order)}
\label{Sec:ODE}For the potential functions (\ref{Eq:HOI})
representing some harmonic-oscillator interaction, all radial
integral eigenvalue equations derived in Sec.~\ref{Sec:REE}
simplify to second-order homogeneous~linear differential
equations; these read, for kernels of Lorentz-scalar Dirac
structure $\Gamma\otimes\Gamma=1\otimes1,$
$$\left[2\,E(p)+a\left(\frac{1}{E^2(p)}+\frac{m^2\,(p^2-5\,m^2)}{2\,E^6(p)}
-\frac{2\,m^2\,p}{E^4(p)}\,\frac{{\rm d}}{{\rm
d}p}+\frac{m^2}{E^2(p)}\,D_p^{(0)}\right)\right]\phi(p)=M\,\phi(p)$$
or\begin{equation}\left[2\,E(p)+a\left(\frac{2\,p^2+3\,m^2}{2\,E^4(p)}
+\frac{m^2}{E(p)}\,D_p^{(0)}\,\frac{1}{E(p)}\right)\right]\phi(p)=M\,\phi(p)\
,\label{Eq:ODE-S}\end{equation}for kernels of time-component
Lorentz-vector Dirac structure
$\Gamma\otimes\Gamma=\gamma^0\otimes\gamma^0,$
\begin{equation}\left[2\,E(p)+a\left(\frac{2\,p^2+3\,m^2}{2\,E^4(p)}
-D_p^{(0)}\right)\right]\phi(p)=M\,\phi(p)\
,\label{Eq:ODE-TV}\end{equation}for kernels of Lorentz-vector
Dirac structure $\Gamma\otimes\Gamma=\gamma_\mu\otimes\gamma^\mu,$
$$\left[2\,E(p)-\frac{3\,a\,m^4}{E^6(p)}
-\frac{2\,a\,m^2\,p}{E^4(p)}\,\frac{{\rm d}}{{\rm
d}p}-a\left(2-\frac{m^2}{E^2(p)}\right)D_p^{(0)}\right]\phi(p)=M\,\phi(p)$$
or\begin{equation}\left[2\,E(p)
+a\left(\frac{m^2}{E(p)}\,D_p^{(0)}\,\frac{1}{E(p)}-2\,D_p^{(0)}\right)\right]
\phi(p)=M\,\phi(p)\ ,\label{Eq:ODE-V}\end{equation}for kernels of
Lorentz-pseudoscalar Dirac structure
$\Gamma\otimes\Gamma=\gamma_5\otimes\gamma_5,$
\begin{equation}\left[2\,E(p)+a\,\frac{2\,p^2+3\,m^2}{2\,E^4(p)}\right]
\phi(p)=M\,\phi(p)\ ,\label{Eq:ODE-PS}\end{equation}and, for
kernels of the BJK \cite{BJK73,Gross91} Dirac structure
$\Gamma\otimes\Gamma=\frac{1}{2}
\left(\gamma_\mu\otimes\gamma^\mu+\gamma_5\otimes\gamma_5-1\otimes1\right),$
\begin{equation}\left[2\,E(p)-a\,D_p^{(0)}\right]\phi(p)=M\,\phi(p)\
.\label{Eq:ODE-BJK}\end{equation}Note that, for the Lorentz
pseudoscalar $\Gamma\otimes\Gamma=\gamma_5\otimes\gamma_5,$ the
reduced Salpeter equation~with harmonic-oscillator potential is
represented by a pure multiplication operator. This implies that
the resulting spectrum is purely continuous. That is, there are no
bound states at~all. This fact is presumably not evident from the
general representation of the reduced Salpeter equation with
pseudoscalar Lorentz structure as radial integral equation, as
given in Sec.~\ref{Sec:REE}.

\subsection{Transformation to a (zero-eigenvalue) Schr\"odinger
equation}\label{Sec:TSE}For $m\ne0,$ the ordinary differential
equations that represent the reduced Salpeter equation
(\ref{Eq:RSE}) for a harmonic-oscillator interaction of {\em
Lorentz-scalar\/} or {\em Lorentz-vector\/} Dirac structure do not
(yet) resemble the familiar form of Schr\"odinger eigenvalue
equations. However,~both of them can be easily cast into the form
of the ordinary differential equation of second~order
\begin{equation}\left[-\frac{{\rm d}^2}{{\rm d}p^2}-2\,g(p)\,\frac{{\rm
d}}{{\rm d}p}+h(p)\right]\phi(p)=0\ ,\label{Eq:2ODE}\end{equation}
involving two given functions, $g(p)$ and $h(p).$ Here, similarly
to the r\^ole of the mass~$m$ of the two bound-state constituents
and the coupling $a$ of the harmonic-oscillator interaction, the
bound-state mass eigenvalue, $M,$ enters---in the function $h(p)$
only---as a parameter. Then, by substitution of the amplitudes
$\phi(p)$ by $\phi(p)=f(p)\,\psi(p),$ with the
transforming~function$$f(p)=p\exp\left[-\int{\rm d}p\,g(p)\right]\
,\quad\left[\frac{{\rm d}}{{\rm
d}p}+g(p)\right]f(p)=\frac{f(p)}{p}\ ,$$determined up to an
irrelevant constant, Eq.~(\ref{Eq:2ODE}) becomes the eigenvalue
equation for~$\psi(p)$$$\left[-\frac{{\rm d}^2}{{\rm
d}p^2}-\frac{2}{p}\,\frac{{\rm d}}{{\rm d}p}+U(p)\right]\psi(p)=0\
,$$corresponding to eigenvalue $0,$ of the Schr\"odinger operator
$-D_p^{(0)}+U(p),$ with the potential\begin{equation}U(p)\equiv
h(p)-\frac{1}{f(p)}\,\frac{{\rm d}^2f}{{\rm
d}p^2}(p)-2\,\frac{g(p)}{f(p)}\,\frac{{\rm d}f}{{\rm
d}p}(p)=h(p)+\frac{{\rm d}g}{{\rm d}p}(p)+g^2(p)\
;\label{Eq:EP}\end{equation}here the second equality follows most
easily from the differential equation satisfied by $f(p).$ Let us
apply this transformation to the two cumbersome Lorentz structures
under~concern.\begin{itemize}\item In the case of a {\em
Lorentz-scalar\/} Bethe--Salpeter kernel,
$\Gamma\otimes\Gamma=1\otimes1,$ the
integration~of$$g(p)=\frac{1}{p}-\frac{p}{E^2(p)}$$gives
$f(p)=E(p)$ [as may be guessed from Eq.~(\ref{Eq:ODE-S})] whereas
the potential $U(p)$~reads\begin{equation}U(p)=
\frac{E^2(p)}{a\,m^2}\,[M-2\,E(p)]-\frac{1}{m^2}-\frac{1}{2\,E^2(p)}
\ .\label{Eq:APUS}\end{equation}\item For the case of a {\em
Lorentz-vector\/} interaction kernel,
$\Gamma\otimes\Gamma=\gamma_\mu\otimes\gamma^\mu,$ the
integration~of
$$g(p)=\frac{1}{p}+\frac{m^2\,p}{E^2(p)\,[E^2(p)+p^2]}$$(which can
be performed most conveniently by partial fraction decomposition)
yields$$f(p)=\frac{E(p)}{\sqrt{E^2(p)+p^2}}\ ;$$reading off $h(p)$
and inserting it into the expression (\ref{Eq:EP}) for the
potential $U(p)$~entails
\begin{equation}U(p)=\frac{E^2(p)\,[2\,E(p)-M]}{a\,[E^2(p)+p^2]}
-\frac{2\,m^2\,p^2}{E^2(p)\,[E^2(p)+p^2]^2}\
.\label{Eq:APUV}\end{equation}\end{itemize}With these formulations
of harmonic-oscillator reduced Salpeter equations at our disposal,
we may take advantage, in Sec.~\ref{Sec:SAS}, of some general
results derived for Schr\"odinger operators.

\section{Self-Adjointness of ``Reduced-Salpeter'' Operators}\label{Sec:SA}
The main goal of our investigation is the thorough analysis of the
qualitative features~of~the spectra of the reduced Salpeter
equation (\ref{Eq:RSE}) with harmonic-oscillator interaction of
various Lorentz structures by close inspection of the resulting
differential equations given in Sec.~\ref{Sec:ODE}.

In this respect, the first problem that arises is the question of
the {\em self-adjointness\/}~of~the operators in our
(differential) equations (\ref{Eq:ODE-S}) through
(\ref{Eq:ODE-BJK}). It is straightforward to show~that all these
operators are self-adjoint: multiplication by a real-valued
function clearly defines a self-adjoint operator and by
integration by parts it is easy to convince oneself
that~$D_p^{(0)}$~and$$m^2\left(\frac{1}{E^2(p)}\,D_p^{(0)}
-\frac{2\,p}{E^4(p)}\,\frac{{\rm d}}{{\rm
d}p}\right)=\frac{m^2}{E(p)}\,D_p^{(0)}\,\frac{1}{E(p)}+\frac{3\,m^4}{E^6(p)}$$
or$$\frac{m^2}{E(p)}\,D_p^{(0)}\,\frac{1}{E(p)}
=\frac{m^2}{E^2(p)}\,D_p^{(0)}-\frac{2\,m^2\,p}{E^4(p)}\,\frac{{\rm
d}}{{\rm d}p}-\frac{3\,m^4}{E^6(p)}$$also represent self-adjoint
operators. This implies that the corresponding spectra are real.

The reality of all {\em eigenvalues\/} $M$ can be also inferred
from Eqs.~(18) and (16) of Ref.~\cite{Olsson96} along the lines of
argument presented (for the case of the full Salpeter
equation,~however) in Sec.~II of Ref.~\cite{Lagae92a} or Sec.~II
of Ref.~\cite{Resag94}. From Eq.~(\ref{Eq:RSE}), all Salpeter
amplitudes $\Phi(\mbox{\boldmath{$p$}})$~satisfy
\begin{eqnarray*}M\int\frac{{\rm d}^3p}{(2\pi)^3}\,{\rm
Tr}\left[\Phi^\dag(\mbox{\boldmath{$p$}})\,\Phi(\mbox{\boldmath{$p$}})\right]
&=&\int\frac{{\rm
d}^3p}{(2\pi)^3}\left[E_1(\mbox{\boldmath{$p$}})+E_2(\mbox{\boldmath{$p$}})\right]{\rm
Tr}\left[\Phi^\dag(\mbox{\boldmath{$p$}})\,\Phi(\mbox{\boldmath{$p$}})\right]\\
&+&\int\frac{{\rm d}^3p}{(2\pi)^3}\int\frac{{\rm
d}^3q}{(2\pi)^3}\,\sum_\Gamma
V_\Gamma(\mbox{\boldmath{$p$}},\mbox{\boldmath{$q$}})\,{\rm
Tr}\left[\Phi^\dag(\mbox{\boldmath{$p$}})\,\gamma_0\,\Gamma\,
\Phi(\mbox{\boldmath{$q$}})\,\Gamma\,\gamma_0\right].\end{eqnarray*}In
this relation, both the integral multiplied by $M$ on the
left-hand side, that is, the~``norm''
$$\|\Phi\|^2\equiv\int\frac{{\rm d}^3p}{(2\pi)^3}\,{\rm Tr}
\left[\Phi^\dag(\mbox{\boldmath{$p$}})\,\Phi(\mbox{\boldmath{$p$}})\right]$$of
each Salpeter amplitude $\Phi(\mbox{\boldmath{$p$}})$ emerging
from the {\em reduced\/} Salpeter equation, and the first term on
the right-hand side are certainly real; the second term on the
right-hand side is real provided all Lorentz-scalar potential
functions $V_\Gamma(\mbox{\boldmath{$p$}},\mbox{\boldmath{$q$}})$
satisfy
$V^\ast_\Gamma(\mbox{\boldmath{$q$}},\mbox{\boldmath{$p$}})=
V_\Gamma(\mbox{\boldmath{$p$}},\mbox{\boldmath{$q$}})$ and the
Dirac matrices $\Gamma$ satisfy
$\gamma_0\,\Gamma^\dag\,\gamma_0=\pm\Gamma,$ which implies that
the matrices $\tilde\Gamma\equiv\gamma_0\,\Gamma$ are (anti-)
Hermitian, i.e., $\tilde\Gamma^\dag=\pm\tilde\Gamma.$ In contrast
to the full Salpeter equation, for the reduced~Salpeter equation
the norm $\|\Phi\|^2$ of all nonzero solutions
$\Phi(\mbox{\boldmath{$p$}})$ is definitely nonvanishing.
Therefore, all mass eigenvalues $M$ are guaranteed to be real for
reasonable interaction kernels
$K(\mbox{\boldmath{$p$}},\mbox{\boldmath{$q$}}).$

\section{Spectra and Stability of the Bound States}\label{Sec:SAS}
\subsection{General considerations}\label{Sec:GC}Now let us
investigate in turn the spectra corresponding to the different
Lorentz structures. Our task is considerably facilitated by making
use of a fundamental theorem \cite{RS4} about~the spectra of
Hamiltonians with potentials increasing without bounds: a
Schr\"odinger operator $H\equiv-\Delta+V,$ defined as a sum of
quadratic forms, with locally bounded, positive, infinitely rising
potential $V(x),$ that is, $V(x)\to\infty$ for $|x|\to\infty,$ may
be shown \cite{RS4}, by application of the well-known
minimum--maximum principle \cite{RS4,Weinstein72,Thirring3}, to
have a purely discrete spectrum.

\newpage

In the case of harmonic-oscillator interactions of {\em
Lorentz-scalar\/} or {\em Lorentz-vector\/} Dirac structure, we
did not succeed, for $m\ne0,$ to reformulate the radial eigenvalue
equations (\ref{Eq:ODE-S}) and (\ref{Eq:ODE-V}), which fix the
bound-state mass eigenvalues $M,$ as standard
Schr\"odinger~equations. However, in Subsec.~\ref{Sec:TSE} we
managed to cast such unruly eigenvalue equations into
the~form$$[-D_p^{(0)}+U(p)]\,\psi(p)=0\ ,$$with $U(p)$ given in
terms of well-defined potentials $U_1(p),$ $U_2(p)$ by
$U(p)=U_1(p)+M\,U_2(p).$

Accordingly, let us analyze the Hamiltonian operator
$H_U\equiv-\Delta+U.$ For $U\equiv U_1+M\,U_2$ satisfying the
assumptions of the above ``infinitely-rising-potential theorem,''
the spectrum of this operator $H_U$ is, for any $M,$ entirely
discrete. In other words, it consists exclusively of isolated
eigenvalues $\varepsilon(M)$ of finite multiplicity that depend,
of course, on the parameter $M.$ Every single zero of these
functions $\varepsilon_i(M)$ ($i\in Z$) defines a bound-state mass
eigenvalue $M$ of Eq.~(\ref{Eq:ODE-S}) or Eq.~(\ref{Eq:ODE-V}).
The derivative of each such function $\varepsilon_i(M)$ with
respect to $M$ is given, in accordance with the Hellmann--Feynman
theorem \cite{FHT}, by the expectation~value~over the associated
eigenstate $|i\rangle$ ($\langle i|i\rangle=1$) of the derivative
of this operator $H_U$ with respect~to~$M$:$$\frac{{\rm
d}\varepsilon_i}{{\rm d}M}(M)=\left\langle i\left|\frac{\partial
H_U}{\partial M}\right|i\right\rangle=\langle i|U_2|i\rangle\
.$$Then, {\em if}, for all eigenvalues $\varepsilon_i(M),$ this
derivative is strictly definite, i.e., if for given~$i$~either
$$\frac{{\rm d}\varepsilon_i}{{\rm d}M}(M)>0\quad\forall\
M$$or$$\frac{{\rm d}\varepsilon_i}{{\rm d}M}(M)<0\quad\forall\
M$$holds, the discreteness of the spectrum of the Hamiltonian
$H_U$ for appropriate potentials~$U$ translates into the
discreteness of all eigenvalues $M$ of the bound-state equations
(\ref{Eq:ODE-S}) or~(\ref{Eq:ODE-V}).

\subsection{Lorentz-scalar kernel: $\Gamma\otimes\Gamma=1\otimes1$}For
massless bound-state constituents, that is, for $m=0,$ our
harmonic-oscillator reduced Salpeter equation with Lorentz-scalar
interaction kernel $\Gamma\otimes\Gamma=1\otimes1,$
Eq.~(\ref{Eq:ODE-S}), simplifies~to
$$\left(2\,p+\frac{a}{p^2}\right)\phi(p)=M\,\phi(p)\ .$$This
relation involves a pure multiplication operator, which possesses
a purely continuous spectrum but no eigenvalues at all.
Consequently, the harmonic-oscillator reduced Salpeter equation
with Lorentz-scalar kernel will not describe bound states of
massless constituents.

In the case of nonvanishing masses $m\ne0$ of the bound-state
constituents, the auxiliary potential $U(p)$ of
Eq.~(\ref{Eq:APUS}) satisfies, for negative harmonic-oscillator
coupling $a=-|a|<0,$ all requirements of the
``infinitely-rising-potential theorem.'' The term $U_2$ multiplied
by~$M,$$$U_2(p)= \frac{E^2(p)}{a\,m^2}\ ,$$is, for $a<0,$
obviously negative definite. Hence, all our considerations of
Subsec.~\ref{Sec:GC}~apply:$$\frac{{\rm d}\varepsilon_i}{{\rm
d}M}\le\frac{1}{a}<0\quad\forall\ i\ .$$Thus, for nonzero
bound-state constituents' mass the resulting spectrum is purely
discrete.

\newpage

\begin{figure}[ht]\begin{center}\psfig{figure=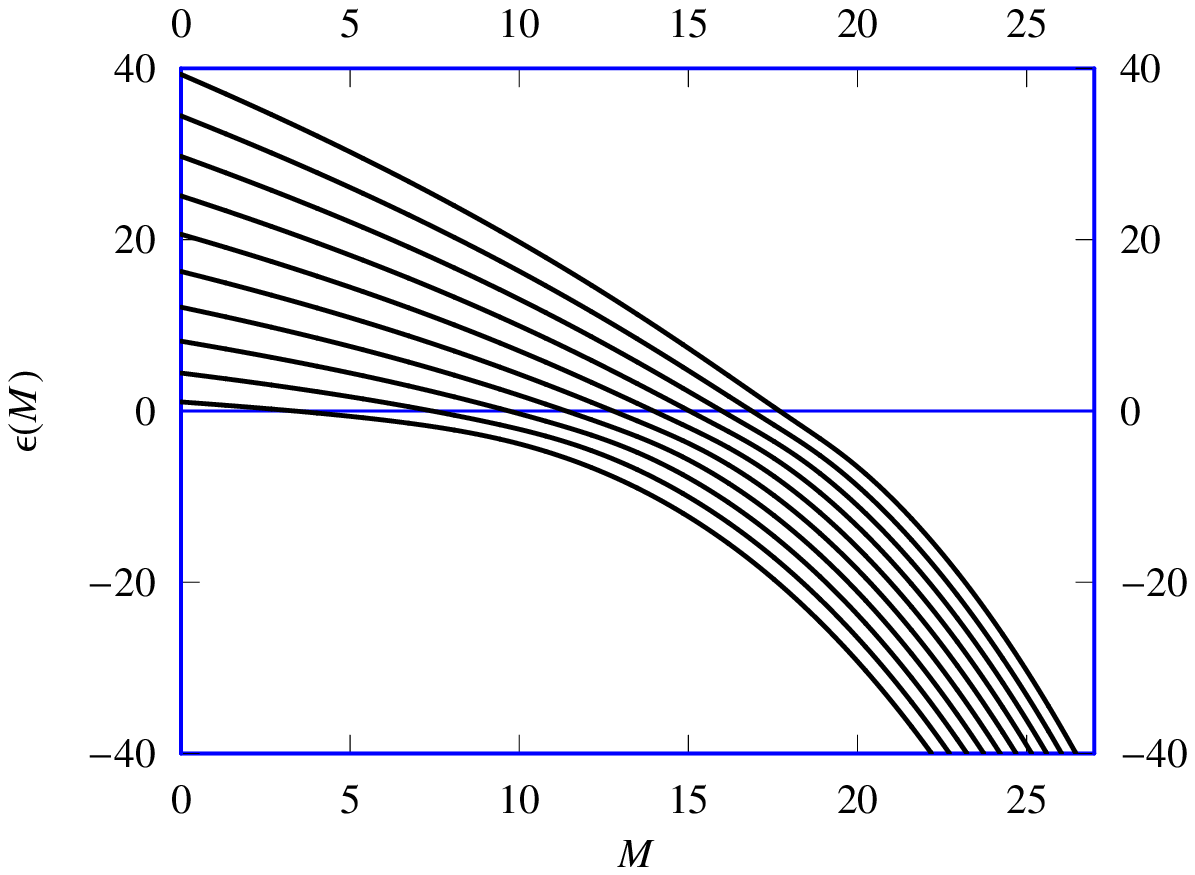,scale=1}
\caption{First, lowest-lying eigenvalues $\varepsilon_i(M),$
$i=0,1,\dots,9,$ of the auxiliary Hamiltonian $H_U\equiv-\Delta+U$
with effective potential $U$ of Eq.~(\ref{Eq:APUS}), corresponding
to the reduced Salpeter equation with harmonic-oscillator
interaction of Lorentz-scalar structure,
$\Gamma\otimes\Gamma=1\otimes1,$~for bound-state constituents'
mass $m=1$ and a ``binding'' coupling $a=-10$ (arbitrary
units).}\label{Fig:AEVLS}\end{center}\end{figure}

Fig.~\ref{Fig:AEVLS} depicts, for a Lorentz-scalar kernel, the
typical behaviour of the discrete auxiliary eigenvalues
$\varepsilon(M)$ for negative harmonic-oscillator couplings $a<0$
producing bound~states.

\subsection{Time-component Lorentz-vector kernel:
$\Gamma\otimes\Gamma=\gamma^0\otimes\gamma^0$}For arbitrary mass
$m\ge0$ of the bound-state constituents, our harmonic-oscillator
reduced Salpeter equation with a time-component Lorentz-vector
interaction kernel $\Gamma\otimes\Gamma=\gamma^0\otimes\gamma^0,$
Eq.~(\ref{Eq:ODE-TV}), yields a Schr\"odinger equation with an
effective potential $V(p)$ in momentum space.

\begin{figure}[p]\begin{center}\begin{tabular}{c}
\psfig{figure=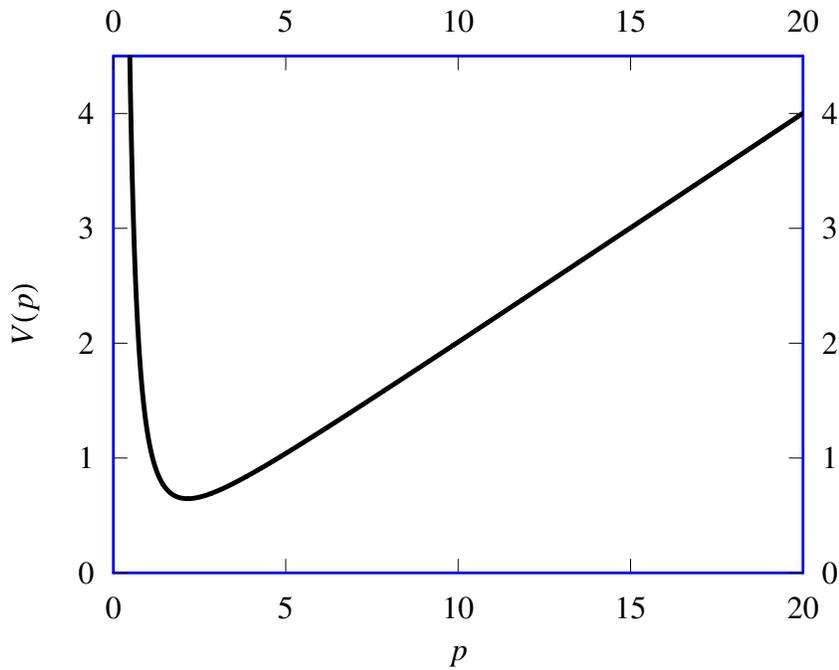,scale=1}\\[0.05ex](a)\end{tabular}\\[1.5ex]
\begin{tabular}{c}\psfig{figure=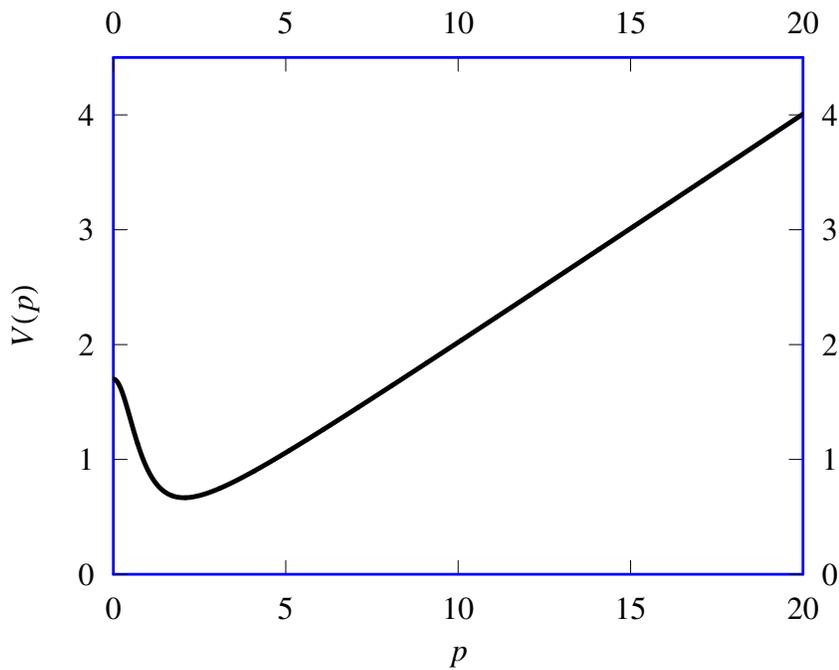,scale=1}\\[0.05ex](b)
\end{tabular}\\[0.5ex]\caption{Qualitative behaviour of the
effective potential, $V(p),$ in the differential-equation
representation (\ref{Eq:ODE-TV}) of the reduced Salpeter equation
with harmonic-oscillator interaction in time-component
Lorentz-vector kernel,
$\Gamma\otimes\Gamma=\gamma^0\otimes\gamma^0,$ for $a=10$ and (a)
vanishing~mass $m=0$ and (b) nonvanishing mass $m=1$ of the
bound-state constituents (arbitrary units).}\label{Fig:VTCLV}
\end{center}\end{figure}

For massless particles, i.e., for $m=0,$ this potential $V(p)$ is
singular at $p=0$ (Fig.~\ref{Fig:VTCLV}):
$$V(p)=\frac{2\,p}{a}+\frac{1}{p^2}\ .$$This potential, however,
is the sum $V(p)=W(p)+1/p^2$ of a linear potential $W(p)=2\,p/a,$
which is positive for a positive slope of this linear rise, that
is, for all $a>0,$ and the singular but positive function $1/p^2.$
Thus, the functions $V(p)$ and $W(p)$ are related by the
inequality $V(p)\ge W(p).$ A suitable combination
\cite{Lucha:Oberwoelz,Lucha:Dubrovnik,Lucha96rcpaubel,Lucha99-1dimsrcp,Lucha04:TWR}
of the minimum--maximum principle \cite{RS4,Weinstein72,Thirring3}
with the resulting operator inequality $H_V\equiv-\Delta+V\ge
H_W\equiv-\Delta+W$ allows to show that any discrete eigenvalue of
$H_V$ is bounded from below by a corresponding eigenvalue of
$H_W.$ Since, by the ``infinitely-rising-potential theorem,'' the
spectrum of $H_W$ is purely discrete---see our detailed discussion
in Subsec.~\ref{Subsec:SAS-V}---, $H_V$ necessarily has a purely
discrete~spectrum. Of course, the same result is obtained by
suitable generalization of the theorem of Ref.~\cite{RS4}.

For nonvanishing mass of the bound-state constituents, $m\ne0,$
the effective potential is fully compatible with the assumptions
of the ``infinitely-rising-potential theorem''
(Fig.~\ref{Fig:VTCLV}):
$$V(p)=\frac{2\,E(p)}{a}+\frac{2\,p^2+3\,m^2}{2\,E^4(p)}\quad
(a>0)\ .$$Here, too, our inevitable conclusion is that the
emerging spectrum must be purely discrete.

\subsection{Lorentz-vector kernel:
$\Gamma\otimes\Gamma=\gamma_\mu\otimes\gamma^\mu$}\label{Subsec:SAS-V}For
massless bound-state constituents, that is, for $m=0,$ our
harmonic-oscillator reduced Salpeter equation with Lorentz-vector
interaction kernel
$\Gamma\otimes\Gamma=\gamma_\mu\otimes\gamma^\mu,$
Eq.~(\ref{Eq:ODE-V}),
becomes$$2\left(p-a\,D_p^{(0)}\right)\phi(p)=M\,\phi(p)\ .$$Of
course, this is nothing but the standard nonrelativistic
Schr\"odinger equation with~linear potential, $V(p)=p/a,$ which is
equivalent to the Airy differential equation \cite{Abramowitz}. To
see~this, introduce a reduced radial wave function
$\varphi(p)\equiv p\,\phi(p)$ and perform the change of variables
$$z=\frac{1}{a^{1/3}}\left(p-\frac{M}{2}\right)$$to a convenient
dimensionless variable, $z,$ in order to arrive at the Airy
differential equation$$\frac{{\rm d}}{{\rm d}z}w(z)=z\,w(z)$$for
$w(z)=\varphi(p).$ Thus, for $m=0$ and a positive slope of this
linear potential, i.e.,~for~$a>0,$ the
``infinitely-rising-potential theorem'' guarantees the pure
discreteness of this spectrum.

\begin{figure}[ht]\begin{center}\psfig{figure=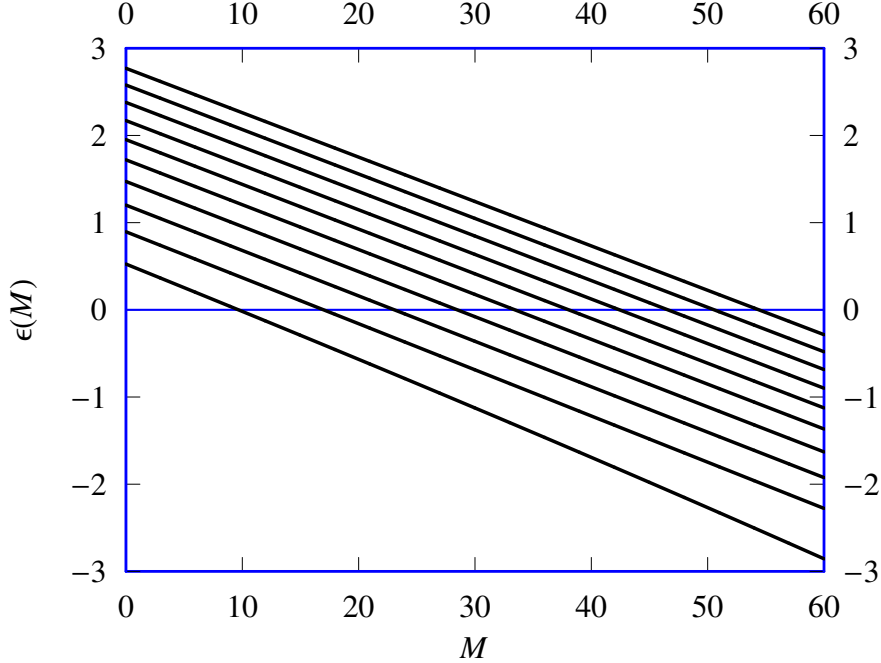,scale=1}
\caption{First, lowest-lying eigenvalues $\varepsilon_i(M),$
$i=0,1,\dots,9,$ of the auxiliary Hamiltonian $H_U\equiv-\Delta+U$
with effective potential $U$ of Eq.~(\ref{Eq:APUV}), corresponding
to the reduced Salpeter equation with a harmonic-oscillator
interaction of Lorentz-vector structure
$\Gamma\otimes\Gamma=\gamma_\mu\otimes\gamma^\mu,$ for bound-state
constituents' mass $m=1$ and ``binding'' coupling $a=10$
(arbitrary units).}\label{Fig:AEVLV}\end{center}\end{figure}

For nonvanishing bound-state constituents' masses, i.e., $m\ne0,$
we again have to~invoke the transformation to an auxiliary
Hamiltonian $H_U$ performed in Subsec.~\ref{Sec:TSE}. The
resulting effective potential $U(p)$ of Eq.~(\ref{Eq:APUV}) is,
for positive harmonic-oscillator couplings $a>0,$~fully compatible
with all the needs of the ``infinitely-rising-potential theorem.''
Its ``$M$-part''~$U_2,$ $$U_2(p)=-\frac{E^2(p)}{a\,[E^2(p)+p^2]}\
,$$is, for all $a>0,$ negative definite: $U_2<0.$ Following our
line of argument of Subsec.~\ref{Sec:GC}, one finds that, also for
nonzero bound-state constituents' mass, the spectrum is purely
discrete. Without surprise, the behaviour of the discrete
auxiliary eigenvalues $\varepsilon(M)$ for $a>0$~is~for a
Lorentz-vector kernel rather similar to that of their
Lorentz-scalar counterparts (cf.~Fig.~\ref{Fig:AEVLV}).

\subsection{Lorentz-pseudoscalar kernel:
$\Gamma\otimes\Gamma=\gamma_5\otimes\gamma_5$}Already in
Sec.~\ref{Sec:ODE}, our inspection of the differential equations
revealed that for any kernel of Lorentz-pseudoscalar type the
spectrum is purely continuous: there is no stability problem.

\subsection{BJK Lorentz structure: $\Gamma\otimes\Gamma=\frac{1}{2}
\left(\gamma_\mu\otimes\gamma^\mu+\gamma_5\otimes\gamma_5-1\otimes1\right)$}For
all $m\ge0,$ a harmonic-oscillator reduced Salpeter equation
(\ref{Eq:ODE-BJK}) for a kernel of the BJK Lorentz structure
$\Gamma\otimes\Gamma=\frac{1}{2}
\left(\gamma_\mu\otimes\gamma^\mu+\gamma_5\otimes\gamma_5-1\otimes1\right)$
is just the eigenvalue equation~of a Schr\"odinger operator with
potential $V(p)=2\,E(p)/a.$ From the ``infinitely-rising-potential
theorem'' we safely conclude that the spectrum of this operator is
purely discrete for $a>0.$

\section{Digression: Investigation of Full Salpeter Equation}
\label{Sec:GFSE}It goes without saying that analogous spectral
analyses \cite{StabOSS-QCD@Work07,Lucha07:SSSECI-Hadron07} can be
envisaged \cite{HOSE} for the Salpeter equation (\ref{Eq:SE}).
There, however, we expect to face more severe problems because any
solution $\Phi(\mbox{\boldmath{$p$}})$ of the Salpeter equation
(\ref{Eq:SE}) involves more than one independent component. Thus a
Salpeter equation with harmonic-oscillator interaction yields a
{\em system\/} of more than one second-order differential equation
or, equivalently, a higher-order differential equation.

Hence, let us recall some general features of the eigenvalues of
the full Salpeter equation (\ref{Eq:SE}), regarded as an
eigenvalue equation for the Salpeter amplitude
$\Phi(\mbox{\boldmath{$p$}});$ these observations emerge either
from the relationship, that is, more precisely, the equivalence,
of the Salpeter equation to the well-known random phase
approximation (RPA) familiar from the study of collective
excitations in nuclear physics or from the inspection of the
Salpeter equation (\ref{Eq:SE}). The normalization condition for
Bethe--Salpeter amplitudes arising as a consequence~of~the
inhomogeneous Bethe--Salpeter equation \cite{BSE} motivates the
introduction of a ``norm'' $\|\Phi\|$~of the Salpeter amplitude
\cite{LeYaouanc85,Lagae92a,Resag94,Olsson95 }, given by (cf.\
Eq.~(2.9) of Ref.~\cite{Lagae92a} or Eq.~(9) of
Ref.~\cite{Olsson95})\begin{eqnarray*}\|\Phi\|^2&\equiv&
\frac{1}{2}\int\frac{{\rm d}^3p}{(2\pi)^3}\,{\rm
Tr}\left[\Phi^\dag(\mbox{\boldmath{$p$}})\left(
\frac{H_1(\mbox{\boldmath{$p$}})}{E_1(\mbox{\boldmath{$p$}})}\,
\Phi(\mbox{\boldmath{$p$}})-\Phi(\mbox{\boldmath{$p$}})\,
\frac{H_2(-\mbox{\boldmath{$p$}})}{E_2(\mbox{\boldmath{$p$}})}
\right)\right]\\&=&\int\frac{{\rm d}^3p}{(2\pi)^3}\,{\rm
Tr}\left[\Phi^\dag(\mbox{\boldmath{$p$}})\,
\frac{H_1(\mbox{\boldmath{$p$}})}{E_1(\mbox{\boldmath{$p$}})}\,
\Phi(\mbox{\boldmath{$p$}})\right].\end{eqnarray*}The two
expressions on the right-hand side of the above definition are, of
course, equivalent by virtue of the constraint (\ref{Eq:SEC})
satisfied by any solution of the Salpeter equation. Because of the
Hermiticity
$H_i^\dag(\mbox{\boldmath{$p$}})=H_i(\mbox{\boldmath{$p$}})$ of
the single-particle Dirac Hamiltonian
$H_i(\mbox{\boldmath{$p$}}),$ $i=1,2,$ the {\em square\/}
$\|\Phi\|^2$ of this ``norm'' is certainly real. It is, however,
{\em not\/} necessarily positive definite. (Hence, the Salpeter
``norm'' $\|\Phi\|$ is not necessarily real, whence the quotation
marks.) The spectrum of Salpeter's equation then exhibits the
following characteristics
\cite{Long84,Lagae92a,Resag93,Resag94,Olsson95,Parramore95,Parramore96}.
\begin{itemize}\item The Salpeter equation can be shown to be
of the same algebraic structure as the RPA equation
\cite{Long84,Resag93,Resag94,Parramore95,Parramore96}. In other
words, Salpeter's equation and the RPA equation~prove to be
equivalent \cite{Resag93,Parramore95,Parramore96}. Now, any RPA
equation can always be rewritten in the form of a self-adjoint
eigenvalue equation for the {\em square\/} of the energy
\cite{Parramore95,Parramore96}. Accordingly, the square of the
energy is guaranteed to be real, which implies that all
eigenvalues~of such RPA equation are either real or purely
imaginary \cite{Long84}. For the Salpeter equation (\ref{Eq:SE})
this means that the squares $P_0^2$ of the energy eigenvalues
$P_0$ of a given bound~state are real or (in the
center-of-momentum or ``rest'' frame of the bound state,
defined~by $\mbox{\boldmath{$P$}}=\mbox{\boldmath{$p$}}_1
+\mbox{\boldmath{$p$}}_2=\mbox{\boldmath{$0$}}$) that the square
$M^2$ of any bound-state mass $M$ is real, respectively.\item In
the center-of-momentum frame of the bound state, any solution
$\Phi(\mbox{\boldmath{$p$}})$ of Salpeter's equation (\ref{Eq:SE})
is subject to the relation (cf.\ Eq.~(2.17) of
Ref.~\cite{Lagae92a} or Eq.~(11) of Ref.~\cite{Olsson95})
\begin{eqnarray}M\,\|\Phi\|^2&=&\int\frac{{\rm
d}^3p}{(2\pi)^3}\left[E_1(\mbox{\boldmath{$p$}})
+E_2(\mbox{\boldmath{$p$}})\right]{\rm
Tr}\left[\Phi^\dag(\mbox{\boldmath{$p$}})\,
\Phi(\mbox{\boldmath{$p$}})\right]\nonumber\\&+&\int\frac{{\rm
d}^3p}{(2\pi)^3}\int\frac{{\rm d}^3q}{(2\pi)^3}\,{\rm
Tr}\left[\Phi^\dag(\mbox{\boldmath{$p$}})\,\gamma_0\,
[K(\mbox{\boldmath{$p$}},\mbox{\boldmath{$q$}})\,
\Phi(\mbox{\boldmath{$q$}})]\,\gamma_0\right]\nonumber\\
&=&\int\frac{{\rm d}^3p}{(2\pi)^3}\left[E_1(\mbox{\boldmath{$p$}})
+E_2(\mbox{\boldmath{$p$}})\right]{\rm
Tr}\left[\Phi^\dag(\mbox{\boldmath{$p$}})\,
\Phi(\mbox{\boldmath{$p$}})\right]\nonumber\\&+&\int\frac{{\rm
d}^3p}{(2\pi)^3}\int\frac{{\rm d}^3q}{(2\pi)^3}\,\sum_\Gamma
V_\Gamma(\mbox{\boldmath{$p$}},\mbox{\boldmath{$q$}})\,{\rm
Tr}\left[\Phi^\dag(\mbox{\boldmath{$p$}})\,\gamma_0\,\Gamma\,
\Phi(\mbox{\boldmath{$q$}})\,\Gamma\,\gamma_0\right].
\label{Eq:SEVSP}\end{eqnarray}For all interaction kernels such
that $\int{\rm d}^3p\int{\rm d}^3q\,{\rm
Tr}\left[\Phi^\dag(\mbox{\boldmath{$p$}})\,\gamma_0\,
[K(\mbox{\boldmath{$p$}},\mbox{\boldmath{$q$}})\,
\Phi(\mbox{\boldmath{$q$}})]\,\gamma_0\right]$ is
real,\begin{eqnarray*}&&\int\frac{{\rm
d}^3p}{(2\pi)^3}\int\frac{{\rm d}^3q}{(2\pi)^3}\left({\rm
Tr}\left[\Phi^\dag(\mbox{\boldmath{$p$}})\,\gamma_0\,
[K(\mbox{\boldmath{$p$}},\mbox{\boldmath{$q$}})\,
\Phi(\mbox{\boldmath{$q$}})]\,\gamma_0\right]\right)^\ast
\\&&=\int\frac{{\rm
d}^3p}{(2\pi)^3}\int\frac{{\rm d}^3q}{(2\pi)^3}\,{\rm
Tr}\left[\Phi^\dag(\mbox{\boldmath{$p$}})\,\gamma_0\,
[K(\mbox{\boldmath{$p$}},\mbox{\boldmath{$q$}})\,
\Phi(\mbox{\boldmath{$q$}})]\,\gamma_0\right],\end{eqnarray*}the
right-hand side of Eq.~(\ref{Eq:SEVSP}) is, of course, real. In
this case, remembering the~reality of the square of the Salpeter
``norm'', the mass eigenvalues $M$ of all bound states~with
nonvanishing ``norm'' $\|\Phi\|$ of the associated Salpeter
amplitudes $\Phi,$ that is, $\|\Phi\|^2\ne0,$ are real: $M^\ast=M$
\cite{Lagae92a,Resag94}. The needed reality of the interaction
term in Eq.~(\ref{Eq:SEVSP}) holds, for instance, for all
Bethe--Salpeter kernels which are sums of terms of Dirac structure
$\Gamma\otimes\Gamma$ such that the Dirac matrices
$\tilde\Gamma\equiv\gamma_0\,\Gamma$ are (anti-) Hermitian,
$\tilde\Gamma^\dag=\pm\tilde\Gamma,$ with the associated
Lorentz-scalar potential function
$V_\Gamma(\mbox{\boldmath{$p$}},\mbox{\boldmath{$q$}})$ satisfying
$V^\ast_\Gamma(\mbox{\boldmath{$q$}},\mbox{\boldmath{$p$}})=
V_\Gamma(\mbox{\boldmath{$p$}},\mbox{\boldmath{$q$}}).$\footnote{Thus,
(at least) for the present class of interactions any solution of
Salpeter's equation (\ref{Eq:SE}) with nonreal and therefore
purely imaginary mass eigenvalue $M^\ast\ne M=\pm{\rm i}\,|M|$
must have vanishing ``norm''~$\|\Phi\|^2=0.$}\item The comparison
of Salpeter's equation (\ref{Eq:SE}) with its Hermitian conjugate
(sandwiched between $\gamma_0$ from the left and $\gamma_0$ from
the right) reveals \cite{Lagae92a,Resag94} that the solutions of
this eigenvalue equation always occur in pairs of the form
$(\Phi,M)$ and $(\Psi,-M^\ast),$ with the Salpeter amplitude
$\Psi(\mbox{\boldmath{$p$}})$ related to the Hermitian conjugate
of its ``partner solution'' $\Phi(\mbox{\boldmath{$p$}}),$
provided the interaction kernel satisfies
$[K(\mbox{\boldmath{$p$}},\mbox{\boldmath{$q$}})\,
\Phi(\mbox{\boldmath{$q$}})]^\dag=\gamma_0\,
[K(\mbox{\boldmath{$p$}},\mbox{\boldmath{$q$}})\,
\Psi(\mbox{\boldmath{$q$}})]\,\gamma_0;$ the squares of the
``norms'' of conjugate solutions have opposite sign:
$\|\Psi\|^2=-\|\Phi\|^2.$ For the case of equal masses $m_1=m_2$
of the bound-state constituents \cite{Lagae92a}, the relation
between $\Phi$ and $\Psi$ is particularly simple, viz.,
$\Psi(\mbox{\boldmath{$p$}})\equiv
\gamma_0\,\Phi^\dag(\mbox{\boldmath{$p$}})\,\gamma_0.$ For the
general case of unequal masses $m_1\ne m_2$ of the bound-state
constituents \cite{Resag94} the relation of $\Phi$ and $\Psi$
involves, in addition, the charge-conjugation matrix $C$ defined,
e.g., by $C\equiv{\rm i}\,\gamma^2\,\gamma^0.$ This doubling of
the eigenvalues to pairs of opposite sign is well known from
solutions to the RPA equation \cite{Long84,Resag93,Resag94}. The
requirement on the interaction kernel holds, for example, for
every Bethe--Salpeter kernel which is a sum of terms each of which
is the product of a real Lorentz-scalar interaction function
$V_\Gamma(\mbox{\boldmath{$p$}},\mbox{\boldmath{$q$}})
=V^\ast_\Gamma(\mbox{\boldmath{$p$}},\mbox{\boldmath{$q$}})$ and a
Dirac structure $\Gamma\otimes\Gamma$ such that all Dirac matrices
$\tilde\Gamma\equiv\gamma_0\,\Gamma$ are (anti-) Hermitian,
$\tilde\Gamma^\dag=\pm\tilde\Gamma.$ Furthermore, the relation
between $\Phi(\mbox{\boldmath{$p$}})$ and
$\Psi(\mbox{\boldmath{$p$}})$ holding within the particular class
of interactions defined by the requirement
$[K(\mbox{\boldmath{$p$}},\mbox{\boldmath{$q$}})\,
\Phi(\mbox{\boldmath{$q$}})]^\dag=\gamma_0\,
[K(\mbox{\boldmath{$p$}},\mbox{\boldmath{$q$}})\,
\Psi(\mbox{\boldmath{$q$}})]\,\gamma_0$ also entails that a
Salpeter amplitude $\Phi$ solving the Salpeter equation for a
nondegenerate zero mass eigenvalue $M=0$ necessarily has vanishing
Salpeter ``norm:'' $\|\Phi\|=0$ \cite{Resag94}.\end{itemize}In
particular, any momentum-space interaction function
$V_\Gamma(\mbox{\boldmath{$p$}},\mbox{\boldmath{$q$}})$ which can
be represented as the Fourier transform
$V_\Gamma(\mbox{\boldmath{$p$}},\mbox{\boldmath{$q$}})
=V_\Gamma(\mbox{\boldmath{$p$}}-\mbox{\boldmath{$q$}})=\int{\rm
d}^3p\exp[-{\rm
i}\,(\mbox{\boldmath{$p$}}-\mbox{\boldmath{$q$}})\cdot\mbox{\boldmath{$x$}}]
\,V_\Gamma(r)$ of some (real) central potential
$V_\Gamma(r)=V^\ast_\Gamma(r)$ in configuration space experienced
by the two bound-state constituents at distance
$r\equiv|\mbox{\boldmath{$x$}}|$ satisfies both
$V^\ast_\Gamma(\mbox{\boldmath{$p$}},\mbox{\boldmath{$q$}})=
V_\Gamma(\mbox{\boldmath{$p$}},\mbox{\boldmath{$q$}})$ and
$V_\Gamma(\mbox{\boldmath{$q$}},\mbox{\boldmath{$p$}})=
V_\Gamma(\mbox{\boldmath{$p$}},\mbox{\boldmath{$q$}}).$ Thus, for
Dirac matrices $\Gamma$ in the kernel
$K(\mbox{\boldmath{$p$}},\mbox{\boldmath{$q$}})=\sum_\Gamma
V_\Gamma(\mbox{\boldmath{$p$}},\mbox{\boldmath{$q$}})\,\Gamma\otimes\Gamma$
obeying $\gamma_0\,\Gamma^\dag\,\gamma_0=\pm\Gamma$ both
conditions on the kernel
$K(\mbox{\boldmath{$p$}},\mbox{\boldmath{$q$}})$ required for the
above-mentioned spectral properties to hold are simultaneously
fulfilled. In this case, the spectrum of mass eigenvalues $M$ of
the Salpeter equation (\ref{Eq:SE}) can only involve, in the
complex-$M$ plane, pairs $(M,-M)$~of opposite sign on the real
axis (excluding the origin), with the associated Salpeter
amplitudes having Salpeter ``norms'' squared of opposite sign,
and/or points $M=-M^\ast$ on the imaginary axis (including the
origin), with corresponding Salpeter solutions of vanishing
Salpeter~``norm.''

Merely for illustrative purposes let us now generalize, for a few
examples, the treatment applied to the reduced Salpeter equation
(\ref{Eq:RSE}) in Sec.~\ref{Sec:PSBS} through Sec.~\ref{Sec:ODE}
to the case of the (full) Salpeter equation (\ref{Eq:SE}). As
recalled in Sec.~\ref{Sec:PSBS}, Salpeter amplitudes
$\Phi(\mbox{\boldmath{$p$}})$ for bound~states with spectroscopic
label $n\,^1J_J$ and, therefore, all Salpeter amplitudes
$\Phi(\mbox{\boldmath{$p$}})$ for bound states with
spin-parity-charge conjugation assignment $J^{PC}=0^{-+}$ consist
of two independent Salpeter components
$\phi_1(\mbox{\boldmath{$p$}})$ and
$\phi_2(\mbox{\boldmath{$p$}}).$ The corresponding Salpeter
equation will therefore reduce to a system of two coupled
(integral) equations for two radial wave
functions~$\phi_1(p)$~and~$\phi_2(p).$\footnote{As recalled in
Sec.~\ref{Sec:PSBS}, for pseudoscalar bound states the constraints
on the solutions $\Phi(\mbox{\boldmath{$p$}})$ of the reduced
Salpeter equation (\ref{Eq:RSE}) imply
$\phi_1(\mbox{\boldmath{$p$}})
=\phi_2(\mbox{\boldmath{$p$}})\equiv\phi(\mbox{\boldmath{$p$}}).$
Consequently, for a given Lorentz structure $\Gamma\otimes\Gamma$
of the interaction kernel the radial eigenvalue equation
representing the reduced Salpeter equation~can be derived by
adding the two radial eigenvalue equations related to Salpeter's
equation (\ref{Eq:SE}) after letting~$\phi_1(p)=\phi_2(p).$}

In order to analyze simultaneously Bethe--Salpeter interaction
kernels of Lorentz-scalar and time-component Lorentz-vector Dirac
structure, we introduce a simple sign factor $\sigma$~by
$$\sigma=\left\{\begin{array}{ll}+1\quad\mbox{for
$\Gamma\otimes\Gamma=\gamma^0\otimes\gamma^0$}&\mbox{(time-component
Lorentz-vector interactions)}\ ,\\[1ex]-1\quad\mbox{for
$\Gamma\otimes\Gamma=1\otimes1$}&\mbox{(Lorentz-scalar
interactions)}\ .\end{array}\right.$$(Lorentz-scalar and
time-component Lorentz-vector confining interaction kernels and
their linear combinations attracted, for purely phenomenological
reasons, particular attention in the Bethe--Salpeter descriptions
\cite{Lagae92a,Lagae92b,Parramore95,Parramore96,Olsson95,Olsson96,
Uzzo99,Lucha00:IBSEm=0,Lucha00:IBSE-C4,Lucha00:IBSEnzm,Lucha01:IBSEIAS}
of hadrons as bound states of quarks \cite{Lucha91,Lucha92}.) With
the help of the parameter $\sigma$ and the interaction functions
$V_L(p,q)$ ($L=0,1$) of Sec.~\ref{Sec:REE}, the radial remnants of
the Salpeter equation (\ref{Eq:SE}) for {\em pseudoscalar\/} bound
states are, for both Lorentz-scalar
($\Gamma\otimes\Gamma=1\otimes1$) and time-component
Lorentz-vector
($\Gamma\otimes\Gamma=\gamma^0\otimes\gamma^0$)~kernels,
\begin{eqnarray*}&&2\,E(p)\,\phi_2(p)+\int\limits_0^\infty\frac{{\rm
d}q\,q^2}{(2\pi)^2}\,\sigma\,V_0(p,q)\,\phi_2(q)=M\,\phi_1(p)\ ,\\
&&2\,E(p)\,\phi_1(p)+\int\limits_0^\infty\frac{{\rm
d}q\,q^2}{(2\pi)^2}\left[\sigma\,\frac{m^2}{E(p)\,E(q)}\,V_0(p,q)
+\frac{p\,q}{E(p)\,E(q)}\,V_1(p,q)\right]\phi_1(q)=M\,\phi_2(p)\
.\end{eqnarray*}This set of equations exhibits a very peculiar
structure: For $M=0,$ the equations decouple. For $M\ne0,$ from
one of the equations one of the two independent Salpeter
components,~say $\phi_1(p),$ may be expressed in terms of the
other, $\phi_2(p),$ and inserted into the other equation in order
to reformulate this eigenvalue problem for the square $M^2$ of the
bound-state~mass~$M$:
\begin{eqnarray*}&&4\,E^2(p)\,\phi_2(p)+2\,E(p)\int\limits_0^\infty
\frac{{\rm d}q\,q^2}{(2\pi)^2}\,\sigma\,V_0(p,q)\,\phi_2(q)\\[1ex]
&&+\:\frac{2}{E(p)}\int\limits_0^\infty\frac{{\rm
d}q\,q^2}{(2\pi)^2}
\left[\sigma\,m^2\,V_0(p,q)+p\,q\,V_1(p,q)\right]\phi_2(q)\\[1ex]
&&+\:\int\limits_0^\infty\frac{{\rm d}q\,q^2}{(2\pi)^2}
\left[\sigma\,\frac{m^2}{E(p)\,E(q)}\,V_0(p,q)
+\frac{p\,q}{E(p)\,E(q)}\,V_1(p,q)\right]\int\limits_0^\infty\frac{{\rm
d}k\,k^2}{(2\pi)^2}\,\sigma\,V_0(q,k)\,\phi_2(k)\\[1ex]&&=M^2\,\phi_2(p)
\ .\end{eqnarray*}Following Sec.~\ref{Sec:HOI}, specifying the
interactions to harmonic-oscillator form allows to trade any
integration over the potential functions $V_L(p,q),$ $L=0,1,$ for
the differential operator~$D_p^{(0)}$:
\begin{eqnarray*}&&\left[4\,E^2(p)-\frac{2\,a}{E(p)}
\left(\sigma\,(p^2+2\,m^2)\,D_p^{(0)}+p\,D_p^{(0)}\,p-2\right)\right.\\&&\left.
+\:\frac{a^2}{E(p)}\left(m^2\,D_p^{(0)}+\sigma\,p\,D_p^{(0)}\,p-2\,\sigma\right)
\frac{1}{E(p)}\,D_p^{(0)}\right]\phi_2(p)=M^2\,\phi_2(p)\
.\end{eqnarray*}This is still a homogeneous linear differential
equation but of fourth order. Thus its analysis requires
techniques or tools beyond those familiar from the study of
Schr\"odinger operators.

The Salpeter equation for a Bethe--Salpeter kernel of the BJK
\cite{BJK73,Gross91} Lorentz structure
$\Gamma\otimes\Gamma=\frac{1}{2}
\left(\gamma_\mu\otimes\gamma^\mu+\gamma_5\otimes\gamma_5-1\otimes1\right)$
constitutes an exceptional case; there all interactions enter only
in one of the two relations which form the set of coupled radial
integral equations describing {\em pseudoscalar bound states}, the
other relation being of ``merely'' algebraic nature:
\begin{eqnarray*}&&2\,E(p)\,\phi_2(p)+2\int\limits_0^\infty\frac{{\rm
d}q\,q^2}{(2\pi)^2}\,V_0(p,q)\,\phi_2(q)=M\,\phi_1(p)\
,\\&&2\,E(p)\,\phi_1(p)=M\,\phi_2(p)\ .\end{eqnarray*}Merging
these two relations generates an eigenvalue problem for $M^2,$
posed equivalently~by
\begin{eqnarray*}4\,E^2(p)\,\phi_1(p)+4\int\limits_0^\infty\frac{{\rm
d}q\,q^2}{(2\pi)^2}\,V_0(p,q)\,E(q)\,\phi_1(q)&=&M^2\,\phi_1(p)\
,\\4\,E^2(p)\,\phi_2(p)+4\,E(p)\int\limits_0^\infty\frac{{\rm
d}q\,q^2}{(2\pi)^2}\,V_0(p,q)\,\phi_2(q)&=&M^2\,\phi_2(p)\
.\end{eqnarray*}For the {\em harmonic-oscillator interaction\/}
$V(r)=a\,r^2$ these eigenvalue equations become,~by virtue of
Eq.~(\ref{Eq:HOI}), i.e.,
$q^2\,V_0(p,q)=-(2\pi)^2\,a\,D_p^{(0)}\,\delta(p-q),$ the {\em
ordinary differential equations}\begin{eqnarray*}
4\left[E^2(p)-a\,D_p^{(0)}\,E(p)\right]\phi_1(p)&=&M^2\,\phi_1(p)\
,\\
4\left[E^2(p)-a\,E(p)\,D_p^{(0)}\right]\phi_2(p)&=&M^2\,\phi_2(p)\
.\end{eqnarray*}Applying the substitution
$E(p)\,\phi_1(p)\propto\phi_2(p)$ proves the equivalence of these
formulations. In contrast to the case of an arbitrary interaction
kernel, for the BJK Lorentz structure our harmonic-oscillator
Salpeter problem reduces to a single second-order differential
equation. The differential operator on the left-hand side of this
eigenvalue problem is {\em not\/} self-adjoint. Nevertheless,
according to the general properties of eigenvalues of the Salpeter
equation (\ref{Eq:SE}) summarized at the beginning of this
section, the spectrum of eigenvalue~squares $M^2$ is real.
Recalling, for $a>0,$ our reasoning of Subsec.~\ref{Sec:GC} for
$H_U\equiv-\Delta+U$ with
auxiliary~potential$$U(p)=\frac{E(p)}{a}-\frac{M^2}{4\,a\,E(p)}\
,$$it is trivial to demonstrate, by similar arguments, that the
spectrum of squared eigenvalues $M^2$ and, as a consequence
thereof, the spectrum of {\em mass eigenvalues\/} $M$ are purely
discrete.

\section{Summary, Conclusions, and Outlook}\label{Sec:SCO}The
present investigation has been devoted to an exploration of the
conditions under~which the reduced Salpeter equation with {\em
confining\/} interactions has {\em stable\/} bound-state
solutions. For harmonic-oscillator interactions, the reduced
Salpeter equation becomes in momentum space either an algebraic
relation or a second-order ordinary differential equation
involving the Laplacian $\Delta$ acting on states of angular
momentum $\ell=0$ (i.e., our differential operator $D_p^{(0)}$
introduced in Sec.~\ref{Sec:HOI}). For real harmonic-oscillator
couplings, all corresponding spectra are real. For pseudoscalar
states, where instabilities are expected to appear first, we
showed that, depending on the Lorentz nature of the kernel, the
resulting spectrum is either purely continuous or entirely
discrete, consisting of isolated mass eigenvalues of finite
multiplicity.

As a by-product, the same analysis proves the {\em boundedness
from below\/} of all the~spectra for appropriate choice of the
respective sign of the harmonic-oscillator coupling constant~$a.$
\begin{itemize}\item For interaction kernels of time-component
Lorentz-vector structure
($\Gamma\otimes\Gamma=\gamma^0\otimes\gamma^0$),
Lorentz-pseudoscalar structure
($\Gamma\otimes\Gamma=\gamma_5\otimes\gamma_5$), the (eventually
simple) BJK \cite{BJK73,Gross91} structure,
$\Gamma\otimes\Gamma=\frac{1}{2}
\,(\gamma_\mu\otimes\gamma^\mu+\gamma_5\otimes\gamma_5-1\otimes1)$,
and, if $m=0,$ of Lorentz-scalar structure
($\Gamma\otimes\Gamma=1\otimes1$) or Lorentz-vector structure
($\Gamma\otimes\Gamma=\gamma_\mu\otimes\gamma^\mu$),~each~harmonic-oscillator
ordinary differential equation of Sec.~\ref{Sec:ODE} is, for
arbitrary $a>0,$ the eigenvalue equation of some {\em positive\/}
operator. The entire spectrum of any such operator must be
positive.\item For an interaction kernel of Lorentz-pseudoscalar
structure ($\Gamma\otimes\Gamma=\gamma_5\otimes\gamma_5$), and any
$m\ne0,$ Eq.~(\ref{Eq:ODE-PS}) is, for finite $a<0,$ the
eigenvalue equation of an operator which~is~not positive but {\em
bounded from below}, by $2\,m+3\,a/(2\,m^2),$ as is its
(continuous) spectrum.\item For $m\ne0,$ the ordinary differential
equations, in Sec.~\ref{Sec:ODE}, related to interaction~kernels
of Lorentz-scalar structure, $\Gamma\otimes\Gamma=1\otimes1,$ or
Lorentz-vector structure,
$\Gamma\otimes\Gamma=\gamma_\mu\otimes\gamma^\mu,$ are not of the
(standard) form of eigenvalue equations of some Schr\"odinger
operators. In order to decide on the nature of their spectra we
have to rely on the transformation of Subsec.~\ref{Sec:TSE} to
Schr\"odinger-like auxiliary operators. For $a<0$ in the
Lorentz-scalar case and $a>0$ in the Lorentz-vector case, all
eigenvalues $\varepsilon_i(M),$ $i=0,1,2,\dots,$~of~the latter
operators can be shown to be entirely discrete for all $M,$ and
strictly decreasing functions of $M.$ Because of their strict
decrease, with increasing bound-state mass $M,$ the zeros of the
lowest trajectories $\varepsilon_0(M)$ define not necessarily
positive lower~bounds on the spectra of mass eigenvalues $M,$
proving both spectra to be {\em bounded from below}.\end{itemize}

Altogether this provides a rigorous proof of the stability of the
considered bound~states: their energies form (for couplings of
suitable sign) {\em real discrete\/} spectra {\em bounded from
below}.

Our findings point in the same direction as the observations made
in a purely numerical analysis \cite{Archvadze95} aiming at the
description of quark--antiquark bound states with the help of the
{\em full\/} Salpeter equation (\ref{Eq:SE}). In this
investigation the (instantaneous) interaction between the
bound-state constituents is modeled by the sum of a Coulomb-type
short-range interaction (arising from one-gluon exchange between
quark and antiquark) and a rather sophisticated confining
interaction, which interpolates between a harmonic-oscillator-type
behaviour for small inter-quark distances $r,$ or small masses of
the bound-state constituents, and a linear rise for large
inter-quark distances $r,$ or large masses of the bound-state
constituents. From a point of view of principle it is a pity that
this study has been performed merely at a single point in
free-parameter space, determined (for an equal-weight mixture of
time-component Lorentz-vector and Lorentz-scalar Dirac structures
of this particular confining~interaction) from a fit of the mass
spectrum of experimentally observed mesons. The authors of
Ref.~\cite{Archvadze95} arrive at the following conclusions (for
bound states composed of equal-mass constituents):\begin{itemize}
\item For bound states of heavy-mass constituents, the
solutions prove to be stable for both time-component
Lorentz-vector and Lorentz-scalar Dirac structures of the
confining interaction and, consequently, for any linear
combination of these Lorentz structures.\item For bound states of
light-mass constituents, the solutions are still all stable for a
pure time-component Lorentz-vector Dirac structure of the
confining interaction~but turn out to be mostly unstable for a
pure Lorentz-scalar spin structure of the
confinement.\end{itemize}However, in Ref.~\cite{Archvadze95} it
was also (numerically) shown that neglecting the
``negative-energy'' components of the Salpeter amplitude
$\Phi(\mbox{\boldmath{$p$}})$ by simply imposing onto
$\Phi(\mbox{\boldmath{$p$}})$ the requirement
$\Lambda_1^-(\mbox{\boldmath{$p$}}_1)\,
\Phi(\mbox{\boldmath{$p$}})\,\Lambda_2^+(\mbox{\boldmath{$p$}}_2)=0$
(almost totally) removes the instability of the bound-state energy
levels otherwise showing up in the case of a Lorentz-scalar
confining interaction. According to our discussion of the
component contents of all solutions of both the full and the
reduced Salpeter equations with respect to the energy projectors
$\Lambda_i^\pm(\mbox{\boldmath{$p$}})$ in Sec.~\ref{Sec:RIBSE},
imposition~of such constraint on $\Phi(\mbox{\boldmath{$p$}})$ is
tantamount to the consideration of the {\em reduced\/} Salpeter
equation~(\ref{Eq:RSE}).

Our approach can be clearly generalized \cite{HOSE} to analyze not
only Salpeter's equation~(\ref{Eq:SE}) but, in a similar way, {\em
three-dimensional reductions\/} of the Bethe--Salpeter equation
different from the Salpeter equation; some of these reductions are
reviewed, for instance, in Ref.~\cite{Kopaleishvili01}.

The intention of the present study was to perform, as a purely
theoretical investigation, a {\em rigorous\/} analysis of the
spectral properties of the Salpeter equation (\ref{Eq:RSE}),
irrespective of its four-dimensional origin within the
Bethe--Salpeter formalism. Our pragmatic point of view seems to be
justified by the fact that in the past exactly this bound-state
equation has been applied in numerous treatments of, for instance,
quark--antiquark bound states in quantum chromodynamics (for
details, consult the reviews in Refs.~\cite{Lucha91,Lucha92}). As
already mentioned~in the Introduction, our interest in the problem
of the particular type of instabilities discussed here has been
aroused by their occasional observation
\cite{Olsson95,Archvadze95}, and the attempts to arrive~at full
{\em analytic\/} understanding of them
\cite{Parramore95,Parramore96,Uzzo99}. Nevertheless, we feel
obliged to add a few brief comments on the practical relevance of
this framework and the implications of our~findings.
\begin{enumerate}\item From the very beginning, our considerations
have been confined to three-dimensional reductions of the
Bethe--Salpeter formalism, obtained by assuming all
interactions~to be {\em instantaneous\/} in the center-of-momentum
frame of the two-particle systems under study. The relevance of
all statements about the presence or absence of instabilities in
any three-dimensional reduction (as given by Salpeter's equation)
for the situation~in the four-dimensional Bethe--Salpeter
formalism remains unclear; its discussion~would necessitate a
spectral analysis of its own, which will become much more involved
than our analysis reported above, but is well beyond the scope of
the present investigation.\item As recalled in
Sec.~\ref{Sec:RIBSE}, in addition to some instantaneous
approximation any derivation of, in particular, the Salpeter
equation relies on the assumption of free propagation of all
bound-state constituents, with a constant (i.e.,
momentum-independent) effective or constituent mass which should
adequately parametrize some significant part of the dynamical
self-energy effects. Needless to say, the free-propagator
assumption cannot be compatible with a confining interaction and
is therefore conceptually problematic: In quantum field theory,
the Dyson--Schwinger equations relate the propagators, that is,
the two-point Green functions, to the $n$-point Green functions
which represent the interactions in the Bethe--Salpeter equation.
Therefore, propagators and interactions cannot be chosen
independently from each other. Thus in non-Abelian gauge theories
the simultaneous assumption of free propagation of the bound
particles and confining interactions (as induced by quantum
chromodynamics) are intrinsically inconsistent. Moreover,
phenomena such as the dynamical breakdown of chiral symmetry can
only be taken into account by retaining the exact propagators,
obtained as the solutions of the Dyson--Schwinger equation for the
corresponding two-point Green function of the bound-state
constituents. Their proper incorporation is crucial for the
interpretation of the lowest-lying pseudoscalar quark--antiquark
bound states: as Goldstone bosons. To make a long story short, it
is, of course, very desirable to have also in one's favorite {\em
instantaneous\/} bound-state equation the exact
fermion~propagators at one's~disposal.

One of the attempts to retain in some three-dimensional reduction
(as far as possible) the original wave-function renormalization
and mass functions parametrizing the full propagators of both
fermionic bound-state constituents resulted in the instantaneous
bound-state equation proposed in Ref.~\cite{Lucha05:IBSEWEP}. The
first tentative exploration of some of the implications of this
generalization \cite{Lucha06:GIBSEEQP-C7} of Salpeter's equation
for quark--antiquark bound states may be found in
Ref.~\cite{Lucha05:EQPIBSE}. The latter study utilizes exact
propagators of light quarks, as extracted within a
``renormalization-group-improved rainbow--ladder truncation''
(which scheme has the undeniable advantage to preserve the
axial-vector Ward--Takahashi identity) applied to both the quark
Dyson--Schwinger equation and the meson Bethe--Salpeter equation
\cite{Maris97}. Given the formalism introduced in
Ref.~\cite{Lucha05:IBSEWEP}, our (obvious) next step is a similar
study \cite{Lucha07:REPBSEHI,Lucha07:SSSECI-Hadron07} for our {\em
full-propagator\/} version~of the Salpeter equation
(\ref{Eq:RSE}). Our preliminary results indicate that for
reasonable behaviour of both bound-state constituents' propagator
functions, i.e., nontrivial wave-function renormalization and
dynamical mass, stability (in our sense) can be achieved
\cite{Lucha07:REPBSEHI,Lucha07:SSSECI-Hadron07}. The significance
of these findings for the four-dimensional Bethe--Salpeter
formalism, which includes all the constituents' self-energy
effects, may be judged by future work.\item Of course, one might
argue that a sufficiently precise purely numerical solution of the
bound-state equation in use may suffice to settle the stability
issues once~and~forever. However, as demonstrated, e.g., by the
not really conclusive findings of
Refs.~\cite{Olsson95,Archvadze95} a numerical analysis may give
but a hint of potential problems with unstable solutions. Thus, it
is our conviction that a {\em genuine\/} understanding of the
origin of these troubles and a compelling solution of this problem
may be gained only by some analytic proof.\end{enumerate}Hence, as
our conclusion let us stress once again that, for the solutions of
the {\em instantaneous\/} Bethe--Salpeter formalism, at least, the
kind of analysis proposed above provides a rigorous answer to the
question of stability, defined by our requirement that each bound
state~found belongs to a {\em real\/} and {\em discrete\/}
spectrum that is {\em bounded from below}. Note that the requested
reality of the bound-state masses precludes, for example, any
solutions of tachyonic nature.

\section*{Acknowledgements}Two of us (W.~L.\ and F.~F.~S.) would
like to thank Bernhard Baumgartner, Harald Grosse, and Heide
Narnhofer for many interesting, stimulating, encouraging, or
helpful~discussions.

\end{document}